\documentclass[12pt]{iopart}
\usepackage{graphicx}
\usepackage{amsbsy}
\usepackage{soul,xcolor}
\usepackage[capitalize]{cleveref}

\begin{document}

\title[Near-Axis Magnetic fields for Turbulence Simulations]{The Use of Near-Axis Magnetic Fields for Stellarator Turbulence Simulations}

\author{R. Jorge, M. Landreman}

\address{Institute for Research in Electronics and Applied Physics, University of Maryland, College
Park, MD 20742, USA}
\ead{rjorge@umd.edu}

\begin{abstract}
The design of turbulence optimized stellarators has so far relied on three-dimensional equilibrium codes such as VMEC in order to find the minimum of a given objective function.
In this work, we propose a complimentary approach based on the near-axis expansion to compute the geometry parameters of neoclassicaly optimized stellarators used in turbulence studies.
As shown here, the near-axis expansion can be a reasonable approximation of the geometric parameters relevant for turbulence and stability simulations of the core of existing optimized stellarator designs.
In particular, we examine the geometry coefficients that appear in the gyrokinetic equation, the drift-reduced fluid equations and the ideal ballooning equation.
This approach may allow for the development of new stellarator optimization techniques significantly faster than conventional methods.
\end{abstract}

%
%
%
%
%

\section{Introduction}
\label{sec:intro}

One of the main challenges faced by the nuclear fusion program today is the presence of transport processes that hinder our ability to achieve and sustain ignition.
Magnetic confinement fusion devices with a continuous rotational symmetry, i.e. axisymmetric devices (tokamaks), exhibit charged particle trajectories that conserve their total angular momentum, resulting in constrained orbits that yield low levels of neoclassical transport.
This property is not present, in general, in non-axisymmetric (stellarator) devices, as trapped particles can deviate a substantial amount from their initial toroidal surface and be lost to the wall \cite{Helander2014}.
However, if the magnetic field strength $B=|\mathbf B|$ used to confine the plasma possesses a continuous symmetry (a property called quasisymmetry), angular momentum conservation is restored and trapped particles are confined even if the magnetic field vector $\mathbf B$ does not contain any symmetry \cite{Nuhrenberg1988a,Boozer1995,Garabedian1996}.
Guiding-center orbits and neoclassical transport in quasisymmetric stellarators share many similarities with the tokamak case, as the geometric coefficients needed to compute particle fluxes in the drift-kinetic limit only depend on position through the toroidal flux and $B$ \cite{Boozer1983}.
Quasisymmetry is an example of  omnigenity, i.e., a field where all collisionless trajectories are confined \cite{Landreman2012}.
Another important subset of omnigeneous fields is the set of  fields that are quasi-isodynamic \cite{Helander2009}, where the contours of constant $B$ close poloidally, rather than toroidally or helically.

In order to further improve the plasma performance of magnetic fusion devices, the next step involves the reduction of turbulent transport in both axisymmetric and non-axisymmetric configurations with the ultimate goal of confining a burning plasma and, eventually, reduce the size and cost of future devices.
Such reduction can be achieved by modifications to the magnetic field.
Notable examples include the doubling of the confinement time in the Tokamak à Configuration Variable (TCV) when triangularity is reversed \cite{Camenen2007} and the turbulence optimization of stellarator configurations aided by analytical proxy functions \cite{Proll2015} and by nonlinear gyrokinetic simulations \cite{Xanthopoulos2014}.

The magnetic field configurations used in such studies and, in general, in the design of stellarator configurations, are usually obtained using computationally intensive techniques such as numerical optimization (e.g. by minimizing the symmetry-breaking Fourier modes of $B$ to obtain a quasisymmetric shape).
Besides the high computational cost of such approach, the dimensionality of the solution space using these techniques is unclear and the optimization procedure is highly dependent on the initial conditions used.
In this work, an approach to obtain three-dimensional magnetic fields based on an expansion about the magnetic axis using $\epsilon=r/\mathcal{R}$ as the expansion parameter is considered, where $r$ is the distance from a point in the surface to the axis and $\mathcal{R}$ a scale length representing the major radius of the device (e.g., the inverse of the maximum axis curvature).
Although the construction is based on an expansion on $\epsilon$, it should be able to describe the core region of any quasisymmetric configuration, including those with low aspect ratio.
Such framework, the so-called near-axis expansion, can be expressed using either a direct \cite{Mercier1964,Jorge2020a,Jorge2020} or an inverse \cite{Landreman2018,Landreman2019,Landreman2019a,Landreman2019b,Landreman2020} coordinate approach depending on the coordinate system used.
In order to leverage the findings of Ref. \cite{Landreman2019} where it was shown that experimental quasisymmetric designs are indeed points in the space of the near-axis expansion, an inverse approach using Boozer coordinates is used here.
Compared with typical equilibrium calculation approaches, a quasisymmetric configuration constructed using the near-axis expansion framework yields at least a four order of magnitude reduction of the computational time at each iteration of the optimization procedure \cite{Landreman2019a}.

Using the near-axis expansion, we are able to show that the geometry coefficients relevant for the solution of kinetic and fluid models generated by the near-axis expansion are reasonable approximations of 3D equilibrium calculations that do not make this expansion.
Furthermore, we can specify how many quasisymmetric configurations exist and the number of associated degrees of freedom.
In this case, the solution space can be parametrized by the shape of the magnetic axis and three real numbers: $I_2$, which is proportional to the on-axis toroidal current density, $\sigma(0)$, which vanishes for a stellarator-symmetric design, and $\overline \eta$, which is related to the mean elongation of the poloidal cross-sections and reflects the magnitude of variation of $B$ on a flux surface.
We note that both $I_2$ and $\sigma(0)$ vanish in most experimental designs leading to a single constant $\overline \eta$ in the parameter space of first order quasisymmetric stellarators.

Ultimately, this work aims to show that the near-axis expansion can be used as an effective tool to survey the space of possible magnetic field shapes for stellarator optimization studies that aim for a reduction of turbulent transport.
Although we focus on a comparison of the relevant geometric quantities using quasisymmetric designs that have a reduced number of free parameters, the present derivation includes the expression for the geometric quantities for both general and quasisymmetric stellarator shapes.

This paper is organized as follows.
In \cref{sec:nagk} a set of independent geometric quantities needed for turbulence and stability simulations are identified.
Section \ref{sec:nearaxis} introduces the near-axis expansion formalism for both general and quasisymmetric stellarator configurations.
The geometric quantities identified in \cref{sec:nagk} are then derived analytically in \cref{sec:coefficients} using the near-axis expansion.
The comparison between the expressions derived in \cref{sec:coefficients} and the geometric profiles of existing quasisymmetric designs is performed in \cref{sec:comparison}.
The conclusions follow.

\section{Geometric parameters for kinetic and fluid plasma models}
\label{sec:nagk}

In this section, we review commonly-used models for the simulation of magnetized plasma turbulence and stability in fusion devices and identify a set of independent geometrical quantities needed to specify the magnetic field geometry.
We take the background magnetic field $\mathbf B$ to be a solution of the ideal MHD system of equations \cite{Freidberg2007}.
In the following, we analyze three commonly used models in magnetic confinement studies: gyrokinetics, a kinetic model commonly employed to study plasma turbulence at the core of fusion devices, a drift-reduced two-fluid model, a model commonly employed to study edge turbulence, and the ballooning equation, a model used to assess the stability of plasma perturbations to ballooning modes.

\subsection{Gyrokinetic Modelling}

A well-known framework that evolves an gyroaveraged distribution function $\left<f_s\right>$ is gyrokinetic theory \cite{Catto1978a,Frieman1982,Brizard2007a,Parra2008,Hahm2009,Abel2013a,Frei2020}.
In here, the particle motion is split into a rapid motion about the magnetic field and the movement of this orbit's centre (the so-called guiding-centre).
As an example of a typical form of the gyrokinetic equation used in turbulence codes such as GS2 \cite{Kotschenreuther1995}, \textit{stella} \cite{Barnes2019} and GENE \cite{Jenko2000a}, we mention the lowest-order, electrostatic gyrokinetic equation for the evolution of the perturbed distribution of guiding centres $g_s=\left<\delta f_s\right>$ with $\delta f_s = f_s-F_s$ and $F_s$ the mean distribution (usually taken to be a Maxwellian)
\begin{eqnarray}\fl
    \frac{\partial g_s}{\partial t}+v_\parallel \mathbf b \cdot \nabla z \left(\frac{\partial g_s}{\partial z}+\frac{q_s}{T_s}\frac{\partial \left<\phi\right>}{\partial z}F_s\right)-\frac{\mu}{m_s}\mathbf b \cdot \nabla z \frac{\partial B}{\partial z} \frac{\partial g_s}{\partial v_\parallel}+\mathbf v_{Ms}\cdot\left(\nabla_\perp g_s+\frac{q_s}{T_s}\nabla_\perp \left< \phi \right> F_s\right)\nonumber\\
    +\left<\mathbf v_E\right>\cdot \nabla_\perp g_s + \left< \mathbf v_E\right> \cdot \nabla F_s = \left< C(\delta f_s)\right>.
\label{eq:nlgk}
\end{eqnarray}
In Eq. (\ref{eq:nlgk}), $v_\parallel$ is the parallel speed, $\mathbf b=\mathbf B/B$ is the magnetic field unit vector, $\mu=m_s v_\perp^2/2B$ is the magnetic moment, $\phi$ is the electrostatic potential, $z$ is any coordinate that measures the location along the magnetic field, $T_s$ is the species temperature, $\mathbf v_E$ is the $\mathbf E \times \mathbf B=\mathbf b \times \nabla_\perp \phi/B$ drift velocity, $\nabla_\perp =-\mathbf b \times(\mathbf b \times \nabla )$ is the perpendicular gradient operator and $\mathbf v_{Ms}$ is the sum of the grad-B and curvature drift velocities
\begin{equation}
    \mathbf v_{Ms}=\frac{\mathbf{b}}{\Omega_s}\times\left(\frac{\mu}{m_s}\nabla B + v_\parallel^2 \boldsymbol \kappa\right),
\end{equation}
with $\Omega_s$ the gyrofrequency and $\boldsymbol \kappa = \mathbf b \cdot \nabla \mathbf b$ the magnetic field curvature.
We note that the gyroaveraging operator $\left<\phi\right>$ can be simplified by taking the Fourier transform in two specified directions $x$ and $y$ orthogonal to $z$, yielding $\left<\phi\right>=\int d \mathbf k \exp(i \mathbf k \cdot \mathbf x)J_0(k_\perp \rho)\phi_{\mathbf k}$ with $J_0$ the zeroth-order Bessel function, $\mathbf k_\perp=k_\alpha \nabla \alpha + k_\psi \nabla \psi$ the perpendicular wave-vector with $\psi$ and $\alpha$ the components of the Clebsch representation of $\mathbf B$ defined below, $\rho=v_\perp /\Omega_s$ the Larmor radius and $\phi_{\mathbf k}$ the Fourier transformed electrostatic potential.

We are now in a position to determine the set of independent geometrical quantities needed to solve the gyrokinetic equation.
In the following, we express the magnetic field $\mathbf B$ as
\begin{equation}
    \mathbf B = \nabla \psi \times \nabla \alpha,
\label{eq:clebsch}
\end{equation}
where $\alpha$ is the field line label and $2\pi\psi$ is the magnetic toroidal flux.
As a spatial coordinate system, we take $(\psi,\alpha,z)$, where for the moment $z$ is any quantity independent of $\psi$ and $\alpha$.
Using the Clebsch representation for the magnetic field in Eq. (\ref{eq:clebsch}), we find eight independent geometrical quantities $\mathbf Q$ needed to solve Eq. (\ref{eq:nlgk}), namely
\begin{eqnarray}
    \mathbf Q=&\left\{B, \mathbf b \cdot \nabla z, |\nabla \psi|^2, |\nabla \alpha|^2, \nabla \psi \cdot \nabla \alpha,\right.\nonumber\\
    &\left.(\mathbf b \times \nabla B) \cdot \nabla \alpha,(\mathbf b \times \nabla B) \cdot \nabla \psi, (\mathbf b \times \boldsymbol \kappa)\cdot \nabla \alpha\right\}.
\label{eq:geomQ}
\end{eqnarray}
We note that the quantity $\mathbf b \times \boldsymbol \kappa \cdot \nabla \psi$ is not present in the set $\mathbf Q$ as $\mathbf b \times \boldsymbol \kappa \cdot \nabla \psi=(\mathbf b \times \nabla B \cdot \nabla \psi)/B$ for any solution $\mathbf B$ of the ideal MHD equations.
The differential operators in the gyrokinetic equation can then be cast into partial derivatives $(\psi, \alpha, z, v_\parallel, \mu)$ of $g_s$ with coefficients given by the quantities in Eq. (\ref{eq:geomQ}).

\subsection{Two-fluid Modelling}

Apart from the modelling of turbulent fluctuations using the gyrokinetic equation, there are several fluid models used to predict the performance of tokamak and stellarator plasmas, especially at the edge and scrape-off layer region where the plasma temperature is low enough to consider a fluid approximation.
Such models are usually derived by taking velocity moments of the Boltzmann equation and using a high-collisionality closure scheme in the magnetized limit.
This leads to two-fluid plasma models such as the Braginskii equations \cite{Braginskii1965} or others \cite{Catto2004,Jorge2017}.
The two-fluid plasma equations can be further simplified by taking advantage of the fact that, in some regions of the device (such as the scrape-off layer region), turbulent structures have characteristic spatial scales much larger than the Larmor radius leading to the so-called drift-reduced fluid models.
As an example of the two-fluid modelling used to simulate plasma turbulence in magnetic confinement fusion devices, we examine the equations first derived in Ref. \cite{Zeiler1997}, implemented in the codes GBS \cite{Ricci2012a} and GDB \cite{Zhu2018}. We then identify a set of independent geometric coefficients needed for the drift-reduced fluid model.

We start with the equation describing the evolution of the plasma density $n=n_e=n_i$, the continuity equation
\begin{eqnarray}
    \frac{\partial n}{\partial t}=&-\frac{[\phi,n]}{B}+\frac{2}{e B}\left[n C(T_e)+T_e C(n)-e n C(\phi)\right]\nonumber\\
    &-n(\mathbf b \cdot \nabla z)\frac{\partial V_{\parallel e}}{\partial z}-V_{\parallel e}(\mathbf b \cdot \nabla z)\frac{\partial n}{\partial z}+S_n,
\label{eq:continuity}
\end{eqnarray}
where $e$ is the elementary charge, $V_{\parallel e}$ is the parallel electron fluid velocity and $S_n$ a density source term.
Apart from the $\mathbf b \cdot \nabla z$ coefficient, we identify as geometric parameters of the continuity equation the curvature operator
\begin{equation}
    C(A)=\frac{B}{2}\left(\nabla \times \frac{\mathbf b}{B}\right) \cdot \nabla A,
\label{eq:curv1}
\end{equation}
and the Poisson bracket
\begin{equation}
    [A,B]=\mathbf b \cdot(\nabla_\perp A \times \nabla_\perp B)=B\left(\frac{\partial A}{\partial \psi}\frac{\partial B}{\partial \alpha}-\frac{\partial A}{\partial \psi}\frac{\partial B}{\partial \alpha}\right).
\end{equation}
Using the identity $\nabla \times \mathbf b \cdot \nabla A=(\mathbf b \cdot \nabla z)(\mathbf b \cdot \nabla \mathbf J/B)\partial A/\partial z+\mathbf b \times \boldsymbol \kappa \cdot \nabla A$, we can express $C(A)$ as
\begin{equation}
    C(A)=\frac{1}{2}\left( \mathbf b \times \boldsymbol \kappa \cdot \nabla_\perp A+\frac{\mathbf b \times \nabla B \cdot \nabla_\perp A}{B} \right)+ O\left(\frac{\rho_{th}}{\mathcal R}\right) 
    .
\label{eq:curv2}
\end{equation}
Equation (\ref{eq:curv2}) shows that, up to leading order in the ratio $\rho_{th}/\mathcal R$ with $\rho_{th}=v_{th}/\Omega$ the thermal Larmor radius and $v_{th}=\sqrt{2 T/m}$ the thermal velocity, the curvature operator can be written in terms of the set of coefficients $\mathbf Q$ in Eq. (\ref{eq:geomQ}) and, therefore, all differential operators in the continuity equation, Eq. (\ref{eq:continuity}).

Next, we write the equation describing the evolution of the electrostatic potential $\phi$, the vorticity equation, as
\begin{eqnarray}
    \fl\frac{\partial}{\partial t}\nabla_\perp^2 \phi = -\frac{[\phi,\nabla^2_\perp \phi]}{B}-V_{\parallel i}(\mathbf b \cdot \nabla z)\frac{\partial \phi}{\partial z}\nabla^2_\perp \phi+\frac{2B}{m_i}\left[C(T_e)+\frac{T_e}{n}C(n)\right]+\frac{B}{3 m_i n}C(G_i)\nonumber\\
    +\frac{m_i \Omega_i^2}{e}\left[(\mathbf b \cdot \nabla z)\left(\frac{\partial V_{\parallel i}}{\partial z}-\frac{\partial V_{\parallel e}}{\partial z}\right)+(V_{\parallel i}-V_{\parallel e})(\mathbf b \cdot \nabla z)\frac{1}{n}\frac{\partial n}{\partial z}\right],
\end{eqnarray}
where, in addition to the curvature and Poisson bracket geometric parameters, we find the perpendicular Laplacian operator $\nabla^2_\perp \phi=\nabla_\perp \cdot(\nabla_\perp \phi)$.
Using the expression $\nabla_\perp = \nabla \alpha \partial_\alpha+\nabla \psi \partial_\psi$ for the perpendicular gradient, we write the perpendicular Laplacian operator as
\begin{eqnarray}
    \nabla^2_\perp \phi&=\frac{\partial^2 \phi}{\partial \alpha^2}|\nabla \alpha|^2+\frac{\partial^2 \phi}{\partial \psi^2}|\nabla \psi|^2 + \frac{\partial^2 \phi}{\partial \psi \partial \alpha}2 \nabla \psi \cdot \nabla \alpha+O\left(\frac{\rho_{th}}{\mathcal R}\right).
\label{eq:vorticity}
\end{eqnarray}
The coefficients $|\nabla \alpha|^2, |\nabla \psi|^2$ and $\nabla \psi \cdot \nabla \alpha$ present in Eq. (\ref{eq:vorticity}) are contained in the set $\mathbf Q$.
Regarding the motion along the magnetic field, we look at the ion momentum equation, that reads
\begin{equation}
    \fl m_i n \frac{\partial V_{\parallel i}}{\partial t}=-\frac{m_i n}{B}[\phi,V_{\parallel i}]-m_i n V_{\parallel i}(\mathbf b \cdot \nabla z)\frac{\partial V_{\parallel i}}{\partial z}-\frac{2}{3}(\mathbf b \cdot \nabla z)\frac{\partial G_i}{\partial z}-(\mathbf b \cdot \nabla z)\frac{\partial [n (T_e+T_i)]}{\partial z},
\label{eq:vpari}
\end{equation}
where $G_i$ is the ion stress function
\begin{equation}
    G_i = -3 \eta_{0i}\left[\frac{2}{3}\mathbf b \cdot \nabla z \frac{\partial V_{\parallel i}}{\partial z}+\frac{C(\phi)}{3B}\right],
\end{equation}
with $\eta_{0i}=0.96n T_i \tau_i$ and $\tau_i$ the ion collisional time.
A similar equation is found for $V_{\parallel e}$.
As shown in Eq. (\ref{eq:vpari}), the geometric-dependent terms of the parallel fluid velocity equations are also contained in the set $\mathbf Q$.
Finally, the equation describing the evolution of the plasma temperature, the energy equation, can be written as
\begin{eqnarray}\fl
    \frac{\partial T_e}{\partial t}=-\frac{[\phi,T_e]}{B}-V_{\parallel e}(\mathbf b \cdot \nabla z)\frac{\partial T_e}{\partial z}+\frac{4}{3 eB}\left[\frac{7}{2}T_e C(T_e)+\frac{T_e^2}{n}C(n)-e T_e C(\phi)\right]+S_T\nonumber\\
    \fl +\frac{2}{3e}\left\{T_e(\mathbf b \cdot \nabla z)\left[0.71\frac{\partial V_{\parallel i}}{\partial z}-1.71\frac{\partial V_{\parallel e}}{\partial z}\right]+0.71 T_e(V_{\parallel i}-V_{\parallel e}(\mathbf b \cdot \nabla z)\frac{1}{n}\frac{\partial n}{\partial z}\right\},
\label{eq:te}
\end{eqnarray}
with $S_T$ the plasma heat source.
A similar equation is found for the evolution of the ion temperature.
As Eq. (\ref{eq:te}) depends on geometry only via the parallel gradient, the curvature and the Poisson bracket operators, its geometric terms are also contained in the set $\mathbf Q$.

\subsection{The ballooning equation}

The problem of stability around a closed magnetic field line in the large toroidal number limit can be formulated as a solution to a one-dimensional eigenvalue problem, the ballooning equation.
While the axisymmetric case was first analyzed in the seminal work of Ref. \cite{Connor1978}, the case of general three-dimensional equilibria was derived in Ref. \cite{Correa-Restrepo1978}.
Such criteria is commonly used as an assessment of the overall stability of a stellarator configuration to ballooning modes \cite{Nuhrenberg1986}.
The ballooning equation can be written as \cite{Cooper1984}
\begin{equation}
    (\mathbf B \cdot \nabla)\left[\frac{k_\perp^2}{B^2}(\mathbf B \cdot \nabla) F\right]+\frac{4\pi p'(\psi)}{\iota B^2}\left(\mathbf B \times \mathbf k_\perp \cdot \boldsymbol \kappa\right)F-\frac{\rho k_\perp^2 \gamma^2}{B^2}F=0,
\label{eq:ballmode}
\end{equation}
where $F$ is the eigenfunction of the mode, $\mathbf k_\perp = k_\psi \nabla \psi + k_\alpha \nabla \alpha$ is the perpendicular wave-vector, $\rho$ is the mass density and $\gamma$ is the growth rate of the mode.
An instability is present if $\gamma_{\textnormal{min}}<0$ where $\gamma_{\textnormal{min}}$ is the lowest eigenvalue $\gamma$ of Eq. (\ref{eq:ballmode}).
Noting that $k_\perp^2=k_\alpha^2 |\nabla \alpha|^2+k_\psi^2 |\nabla \psi|^2+2 k_\psi k_\alpha \nabla \psi \cdot \nabla \alpha$, we find that the geometric quantities present in Eq. (\ref{eq:ballmode}) can be written using the parameters in the set $\mathbf Q$ in Eq. (\ref{eq:geomQ}).
We therefore conclude that for the three models outlined above, the quantities needed to fully specify the geometry are contained in the set $\mathbf Q$ in Eq. (\ref{eq:geomQ}).
These are the ones that are analytically derived in the next section using the near-axis expansion formalism and then compared with existing quasisymmetric stellarator designs.

\section{The near-axis expansion formalism}
\label{sec:nearaxis}

In this section we review the near-axis expansion as originally derived by Garren and Boozer \cite{Garren1991a,Garren1991}.
We use a straight field-aligned coordinate system known as Boozer coordinates \cite{Boozer1981} denoted as $(\psi, \theta, \varphi)$, with $\theta$ and $\varphi$ the poloidal and toroidal angles, respectively and $2\pi \psi$ the toroidal magnetic flux.
For convenience, we introduce a helical angle $\vartheta = \theta - N \varphi$ with $N$ a constant integer.
The magnetic field $\mathbf B$ can then be written as
\begin{eqnarray}
\mathbf{B} = \nabla\psi \times\nabla\vartheta + \iota_N \nabla\varphi \times\nabla\psi,
\label{eq:straight_field_lines_h}
\\
 = \beta \nabla\psi + I \nabla\vartheta + (G+NI) \nabla\varphi,
\label{eq:Boozer_h}
\end{eqnarray}
where $\iota_N = \iota - N$ with $\iota$ the rotational transform, and $I, G$ and $\iota$ are constants on $\psi$ surfaces, i.e., $I=I(\psi)$, $G=G(\psi)$ and $\iota = \iota(\psi)$.
Alternatively, the magnetic field can be written in the Clebsch representation of Eq. (\ref{eq:clebsch}) by identifying the field line label $\alpha$ as $\alpha = \vartheta - \iota_N \varphi$.
The vectors $\nabla \psi$, $\nabla \vartheta$ and $\nabla \varphi$ can be related to the position vector $\mathbf r$ using the dual relations
\begin{equation}
    \nabla \psi = \frac{1}{J}\frac{\partial \mathbf r}{\partial \vartheta} \times \frac{\partial \mathbf r}{\partial \varphi},~\frac{\partial \mathbf r}{\partial \psi}=J \nabla \vartheta \times \nabla \varphi,
\label{eq:dualrelations}
\end{equation}
plus all even permutations of $(\psi, \vartheta, \varphi)$, with $J$ the Jacobian defined by
\begin{equation}
    J = \frac{\partial \mathbf r}{\partial \psi} \times \frac{\partial \mathbf r}{\partial \vartheta} \cdot \frac{\partial \mathbf r}{\partial \varphi}= \frac{1}{\nabla \psi \times \nabla \vartheta \cdot \nabla \varphi}=\frac{G+\iota I}{B^2}.
\label{eq:jna}
\end{equation}
For completeness, we write the following expressions for the contravariant components of $\mathbf B$
\begin{eqnarray}
    \mathbf{B} \cdot \nabla \vartheta = \iota_N/J,~\mathbf{B} \cdot \nabla \varphi = 1/J,~\mathbf{B} \cdot \nabla \psi = 0.
\label{eq:idna}
\end{eqnarray}

In the near-axis framework, the position vector $\mathbf{r}$ is written as
\begin{eqnarray}
\label{eq:positionVector}
\mathbf{r}(r,\vartheta,\varphi) = \mathbf{r}_0(\varphi)
+X(r,\vartheta,\varphi) \mathbf{n}(\varphi)
+Y(r,\vartheta,\varphi) \mathbf{b}(\varphi)
+Z(r,\vartheta,\varphi) \mathbf{t}(\varphi),
\end{eqnarray}
where $\mathbf{r}_0(\varphi)$ is the position vector of the magnetic axis and $r$ is a flux surface label defined by $2 \pi \psi = \pi r^2 \bar{B}$ with $\bar{B}$ a constant reference field strength.
The orthonormal vectors $(\mathbf{t},\mathbf{n},\mathbf{b})$ satisfy the Frenet-Serret set of equations
\begin{eqnarray}\fl
\frac{d\varphi}{d\ell}
\frac{d\mathbf{r}_0}{d\varphi} = \mathbf{t}, 
\hspace{0.3in}
\frac{d\varphi}{d\ell}
\frac{d\mathbf{t}}{d\varphi} = \kappa \mathbf{n}, 
\hspace{0.3in}
\frac{d\varphi}{d\ell}
\frac{d\mathbf{n}}{d\varphi} = -\kappa \mathbf{t} + \tau \mathbf{b}, 
\hspace{0.3in}
\frac{d\varphi}{d\ell}
\frac{d\mathbf{b}}{d\varphi} = -\tau \mathbf{n}, 
\label{eq:Frenet}
\end{eqnarray}
with $\mathbf{b}=\mathbf{t}\times\mathbf{n}$, $\ell$ the arclength along the axis, $\kappa=\kappa(\varphi)$ the axis curvature and $\tau=\tau(\varphi)$ the axis torsion.

Near the magnetic axis (in a high-aspect ratio regime where $\epsilon \ll 1$), the Frenet-Serret components of $\mathbf r$ can be expanded as
\begin{eqnarray}
X(r,\vartheta,\varphi)
= r X_1(\vartheta,\varphi) + r^2 X_2(\vartheta,\varphi) + r^3 X_3(\vartheta,\varphi) + \ldots,
\label{eq:radial_expansion}
\end{eqnarray}
with analogous expressions for $Y$ and $Z$. 
Other than $r$, all scale lengths in the system are ordered as $\mathcal{R}$ such that Eq. (\ref{eq:radial_expansion})
represents an expansion in $\epsilon$.
The field strength is expanded similarly but with an $\epsilon^0$ term
\begin{eqnarray}
\label{eq:radial_expansion_B}
B(r,\vartheta,\varphi)
= B_0(\varphi) + r B_1(\vartheta,\varphi) + r^2 B_2(\vartheta,\varphi)+ r^3 B_3(\vartheta,\varphi) + \ldots,
\end{eqnarray}
and $\beta(r,\vartheta,\varphi)$ is expanded in the same way. 
The profile functions
$G(r)$, $I(r)$, $p(r)$, and $\iota_N(r)$ are analytic functions of $\psi=\psi(r^2)$, so their expansions contain only even powers of $r$
\begin{eqnarray}
p(r) = p_0 + r^2 p_2 + r^4 p_4 + \ldots.
\end{eqnarray}
Since $I(r)$ is proportional to the toroidal current inside the surface $r$, then $I_0 = 0$.
From analyticity considerations near the axis (see appendix A of \cite{Landreman2018}), the expansion coefficients have the form
\begin{eqnarray}
\label{eq:poloidal_expansions}
X_1(\vartheta,\varphi) = &X_{1s}(\varphi) \sin(\vartheta) + X_{1c}(\varphi) \cos(\vartheta).
\end{eqnarray}
The parameters $Y$, $Z$, $B$, and $\beta$ are expanded in a similar manner.
For the calculations that follow, it is useful to introduce symbols for the signs of two quantities: $s_G = \textnormal{sgn}(G) = \pm 1$, and $s_\psi = \textnormal{sgn}(\psi) = \textnormal{sgn}(\bar{B}) = \pm 1$.
We note that, due to the definition of $\mathbf B$ in Eqs. (\ref{eq:straight_field_lines_h}) and (\ref{eq:Boozer_h}), each of these signs can be flipped individually by reversing the signs of the poloidal or toroidal angle.
To lowest order in the expansion, $G_0$ and $B_0$ are related via
\begin{equation}
    G_0=s_G B_0 \frac{d\ell}{d\varphi}.
\end{equation}

For the case of quasisymmetry, $B_0$ is constant and the position vector $\mathbf r$ to first order in $r$ can be written as \cite{Landreman2018}
\begin{equation}
    \mathbf r = \mathbf r_0 + r \frac{\overline \eta}{\kappa}\left[\cos \vartheta\mathbf n + \frac{s_\psi s_G\kappa^2}{\overline \eta^2}\left(\sin \vartheta + \sigma \cos \vartheta\right)\mathbf b \right] + O(\epsilon^2),
\label{eq:qspositionvector}
\end{equation}
with $\sigma=\sigma(\varphi)$ a solution of
\begin{equation}
    \frac{d\sigma}{d\varphi}+(\iota_0-N)\left(\frac{\overline \eta^4}{\kappa^4}+1+\sigma^2\right)-\frac{s_G L \overline \eta^2}{\pi \kappa^2}\left(\frac{I_2}{B_0}-s_\psi\tau\right)=0,
\label{eq:sigma}
\end{equation}
where $L$ is the total length of the magnetic axis and the identity $G_0=s_G B_0 L /2\pi$ was used.
As an aside, we note that for a general (non-quasisymmetric) stellarator, an equation analogous to Eq. (\ref{eq:sigma}) holds, which can be found in Ref. \cite{Garren1991} or Eq. (A.26) of Ref. \cite{Landreman2019b}.
The parameter $\overline \eta$ is a constant and reflects the magnitude by which the magnetic field strength $B$ varies on flux surfaces
\begin{equation}
    B=B_0\left[1+r \overline \eta \cos \vartheta+ O(\epsilon^2)\right].
\label{eq:bna}
\end{equation}
Given $\overline \eta$, $I_2$, an axis shape and an initial condition $\sigma(0)$, Eq. (\ref{eq:sigma}) can be solved employing periodic boundary conditions, yielding the function $\sigma$ and the rotational transform on axis $\iota_0$.
The constant $N$ is unique for each axis shape and is given by the total number of poloidal rotations of the normal vector $\mathbf n$ after one toroidal transit.
For details concerning the numerical method used to solve Eq. (\ref{eq:sigma}) see Ref. \cite{Landreman2019a}.

\section{Geometric Quantities}
\label{sec:coefficients}

In this section, we derive the lowest order components in $\epsilon$ of $\mathbf Q$ using the near-axis expansion formalism outlined in the previous section, both for general and quasisymmetric magnetic fields.
The eight independent geometrical quantities $Q_i$ are repeated here for convenience
\begin{eqnarray}
    \mathbf Q=&\left\{B, \mathbf b \cdot \nabla z, |\nabla \psi|^2, |\nabla \alpha|^2, \nabla \psi \cdot \nabla \alpha, \right.\nonumber\\
    &\left.(\mathbf b \times \nabla B) \cdot \nabla \alpha,(\mathbf b \times \nabla B) \cdot \nabla \psi, (\mathbf b \times \boldsymbol \kappa)\cdot \nabla \alpha\right\}.
\label{eq:geomQ1}
\end{eqnarray}

\subsection{General Case}
\label{sec:generalCase}

Starting with $Q_1=B$, we note that, to lowest order, the magnetic field can be written as $\mathbf B \simeq G_0 \nabla \varphi$ which, using the dual relations in Eq. (\ref{eq:dualrelations}), yields
\begin{equation}
    B=B_0 \left[1+ r \kappa (X_{1c}\cos \vartheta  +X_{1s}\sin \vartheta )\right].
\label{eq:g1NA}
\end{equation}
From here onward, we choose $z=\varphi$ (while other choices such as $z=\vartheta$ could be made).
The angle $\vartheta$ can then be related to the coordinate $z$ via $\vartheta=\alpha+\iota_N z$.
The expression for $Q_2=\mathbf b \cdot \nabla z$ can be found by identifying $z=\varphi$ as the magnetic field line following coordinate and taking the lowest order component of Eq. (\ref{eq:idna})
\begin{equation}
    \mathbf b \cdot \nabla z=\frac{B_0}{G_0}\left[1+r \kappa (X_{1c}\cos \vartheta  +X_{1s}\sin \vartheta )\right].
\label{eq:g2NA}
\end{equation}

The next three quantities, $Q_3=|\nabla \psi|^2, Q_4=|\nabla \alpha|^2$ and $Q_5=\nabla \psi \cdot \nabla \alpha$ are found by taking $\nabla \psi = \overline B r \nabla r \simeq r B_0[(Y_{1s}\cos \vartheta - Y_{1c}\sin \vartheta)\mathbf{n}+(X_{1c} \sin \vartheta - X_{1s} \cos \vartheta)\mathbf{b}]$ and $\nabla \alpha = \nabla \vartheta - \iota_N \nabla \varphi-2 \iota_2 r \varphi \nabla r \simeq B_0/(\overline{B}r)[-(Y_{1c}\cos \vartheta +Y_{1s}\sin \vartheta)\mathbf n +(X_{1c} \cos \vartheta + X_{1s} \sin \vartheta)\mathbf b]$, leading to
\begin{eqnarray}
    \fl |\nabla \psi|^2&=r^2 B_0^2 \left[ (X_{1c} \sin \vartheta - X_{1s} \cos \vartheta)^2 +(Y_{1c}\sin \vartheta  - Y_{1s}\cos \vartheta )^2\right] \label{eq:g3NA},\\
    \fl |\nabla \alpha|^2&=\frac{B_0^2}{\overline B^2 r^2} \left[(X_{1c} \sin \vartheta + X_{1s} \cos \vartheta)^2 +(Y_{1c} \cos \vartheta +Y_{1s} \sin \vartheta )^2\right] \label{eq:g4NA},\\
    \fl \nabla \psi \cdot \nabla \alpha&=\frac{B_0^2}{2 \overline B} \left[  \left(X_{1c}^2-X_{1s}^2+Y_{1c}^2-Y_{1s}^2\right)\sin 2 \theta -2 (X_{1c} X_{1s}+Y_{1c} Y_{1s}) \cos 2 \vartheta \right]. \label{eq:g5NA}
\end{eqnarray}
The components of $\mathbf Q$ involving $\nabla B$ read
\begin{eqnarray}
    (\mathbf b \times \nabla B)\cdot \nabla \alpha&= \frac{B_0^3 \kappa}{\overline B r} \left(X_{1c}\cos \vartheta  +X_{1s}\sin \vartheta \right)\label{eq:g6NA},\\
    (\mathbf b \times \nabla B)\cdot \nabla \psi&=r B_0^3 \kappa (X_{1c}\sin \vartheta-X_{1s}\cos \vartheta ). \label{eq:g7NA}
\end{eqnarray}
Finally, the last component of $\mathbf Q$ can be simplified using the identity $\mathbf b \times \boldsymbol \kappa = \mathbf b \times \nabla B/{B}+{\mathbf b \times \nabla p}/{B^2}$.
As the pressure gradient term in the previous identity introduces a component that is higher order in $r$ than the grad-$B$ term, we find
\begin{eqnarray}
    (\mathbf b \times \boldsymbol \kappa)\cdot \nabla \alpha&=\frac{B_0^2 \kappa}{\overline B r} \left(X_{1c}\cos \vartheta  +X_{1s}\sin \vartheta \right) \label{eq:g8NA}.
\end{eqnarray}

\subsection{Quasisymmetry}

The geometric quantities $\mathbf Q$ derived above using the near-axis expansion formalism are now reduced to the quasisymmetric case and the input parameters needed to perform a comparison with existing quasisymmetric designs are identified.
In this case, the position vector in Eq. (\ref{eq:qspositionvector}), is simplified via $X_{1s}=0$, $X_{1c}=\overline \eta/\kappa$, $Y_{1s}=s_G s_\psi \kappa/\overline \eta$ and $Y_{1c}=s_G s_\psi \sigma \kappa/\overline \eta$.
In addition, the lowest order component of $G$ can be written as $G_0=s_G L B_0/2\pi$.
Starting with $Q_1=B$, as shown in Eq. (\ref{eq:bna}), this can be written as
\begin{equation}
    B=B_0(1+r \overline \eta \cos \vartheta).
\label{eq:g1}
\end{equation}
The constants (input parameters) needed to specify $B$ are then $B_0$ (the magnetic field on-axis) and $\overline \eta$.
The parallel gradient parameter $Q_2=\mathbf b \cdot \nabla z$, using Eqs. (\ref{eq:jna}) and (\ref{eq:idna}) together with $G_0=s_G L B_0/2\pi$ and $I_0=0$, yields
\begin{equation}
    \mathbf b \cdot \nabla z=s_G \frac{2\pi}{L}(1+r \overline \eta \cos \vartheta),
\label{eq:g2}
\end{equation}
which adds $L$, hence an axis shape, to the input parameters.

The next three geometric quantities are computed using the quasisymmetric form of the position vector $\mathbf r$ in Eq. (\ref{eq:qspositionvector}) and the dual relations in Eq. (\ref{eq:dualrelations}), yielding
\begin{eqnarray}
    |\nabla \psi|^2&=r^2 \frac{B_0^2}{\overline \eta^2 \kappa^2}\left[\overline \eta^4 \sin ^2\vartheta+\kappa^4 (\cos \vartheta-\sigma\sin \vartheta )^2\right],\label{eq:g3}\\
    |\nabla \alpha|^2&=\frac{1}{r^2\overline \eta^2 \kappa^2}\left[\overline \eta^4 \cos ^2\vartheta+\kappa^4 (\sigma \cos \vartheta +\sin\vartheta)^2\right],\label{eq:g4}\\
    \nabla \psi \cdot \nabla \alpha&=\frac{s_\psi B_0}{2 \overline \eta^2 \kappa^2}\left( \left[\overline \eta^4+\kappa^4 \left(\sigma^2-1\right)\right]\sin 2 \vartheta-2 \kappa^4 \sigma \cos 2 \vartheta\right).\label{eq:g5}
\end{eqnarray}
The function $\sigma=\sigma(\varphi)$ is obtained solving Eq. (\ref{eq:sigma}) given an axis shape and $\overline \eta$. (The configurations used in this study have $\sigma(0)=I_2=0$).
Regarding the $\nabla B$ components of $\mathbf Q$, we find
\begin{eqnarray}
    (\mathbf b \times \nabla B)\cdot \nabla \alpha&=\frac{s_\psi}{r} B_0^2 \overline \eta \cos \vartheta \label{eq:g6},\\
    (\mathbf b \times \nabla B)\cdot \nabla \psi&=r B_0^3 \overline \eta \sin \vartheta, \label{eq:g7}
\end{eqnarray}
while the $\boldsymbol \kappa$ component of $\mathbf Q$ is given by
\begin{eqnarray}
    (\mathbf b \times \boldsymbol \kappa)\cdot \nabla \alpha&=\frac{s_\psi}{r} B_0 \overline \eta \cos \vartheta \label{eq:g8}.
\end{eqnarray}

A key property of quasisymmetric fields is that the guiding center dynamics will be the same as if the particles were in an axisymmetric field \cite{Boozer1983}. 
Therefore, the geometric coefficients present in the guiding center equations of motion, namely the quasisymmetric forms of $B, \mathbf b \cdot \nabla z, (\mathbf b \times \nabla B)\cdot \nabla \alpha, (\mathbf b \times \nabla B)\cdot \nabla \psi, (\mathbf b \times \boldsymbol \kappa)\cdot \nabla \alpha$ and $(\mathbf b \times \boldsymbol \kappa)\cdot \nabla \alpha$ are isomorphic to the ones in a tokamak.
However, the quantities $|\nabla \psi|^2, |\nabla \alpha|^2$ and $\nabla \psi \cdot \nabla \alpha$, which stem from finite Larmor radius contributions to the gyrokinetic and fluid equations, are absent in the guiding-center trajectory equations and may therefore be different when comparing quasisymmetry with axisymmetry.
In a stellarator, these three quantities will have spatial variation reflecting the number of field periods.
Therefore we expect $|\nabla \psi|^2, |\nabla \alpha|^2$ and $\nabla \psi \cdot \nabla \alpha$ to have a faster variation with $z$ in a quasisymmetric stellarator than in a tokamak.

In the system of Eqs. (\ref{eq:g1})-(\ref{eq:g8}), we identify two input parameters needed for this study, namely $B_0, \overline \eta$, together with an axis shape.
The latter is needed in order to compute the axis length $L$, the number of rotations of the normal vector $N$ after one toroidal transit, the curvature $\kappa$ and torsion $\tau$.
While the axis shape can be readily obtained from the considered stellarator designs, $B_0$ and $\overline \eta$ are obtained by seeking a near-axis configuration that closely matches the surface shape and the rotational transform of the inner region of such devices.
A similar study was performed in Ref. \cite{Landreman2019}, where the value of $\overline \eta$ was obtained for several quasisymmetric designs.
Finally, we note that $Q_2$ is a function of $Q_1$ and, to this order, the coefficient $Q_6$ is identical to $Q_8$ (apart from a factor of $B_0)$.
Therefore, in the following, a comparison is performed between the remaining 6 geometrical coefficients in $\mathbf{Q}$, namely $Q_1, Q_3, Q_4, Q_5, Q_6$ and $Q_7$.

\section{Numerical Results}
\label{sec:comparison}

We now compare the geometry coefficients $\mathbf Q$ derived using the near-axis expansion in Section \ref{sec:coefficients} to the ones computed using existing stellarator designs.
For this study, we choose eleven quasisymmetric designs developed by several independent research teams using different optimization codes.
By order of decreasing number of field periods, these are the NZ1988 design of Ref. \cite{Nuhrenberg1988}, the Drevlak design of Ref. \cite{Drevlak2017} recently developed at the Max Planck Institute for Plasma Physics in Greifswald, Germany, HSX \cite{Anderson1995}, the KuQHS48 design of Ref. \cite{Ku2011}, the WISTELL-A configuration of Ref. \cite{Bader2020} recently developed at the University of Wisconsin-Madison, USA, NCSX (configuration LI383,  \cite{Zarnstorff2001}), ARIES-CS (configuration N3ARE,  \cite{Najmabadi2008}), the QAS2 configuration of Ref. \cite{Garabedian2008} with a vanishing on-axis current, ESTELL \cite{Drevlak2013}, CFQS \cite{Shimizu2018a} and the Henneberg design of Ref. \cite{Henneberg2019}.
The properties of each stellarator design are listed in \cref{tab:configurations}.
A magnetic field line with $\alpha=0$ is used.

\begin{table}
\begin{tabular}{@{}llllllll}
&&Field&Plasma&$\iota$ from&&\\
&Aspect&periods&pressure&optimized&$\iota$ from&Best-fit \\
Configuration&ratio $A$&$n_{fp}$&$\beta$&configuration&construction&$\overline \eta$ [m$^{-1}$]&$B_0$ [T]\\
\hline
NZ1988 &12&6& 0\%&1.42 & 1.42 & 0.157&0.205\\
Drevlak&8.6&5&4\%&1.50&1.50&0.0899&3.97\\
HSX&10&4&0\%&1.05&1.06&1.28&1.00\\
KuQHS48&8.1&4&4\%&1.29&1.27&0.147&1.20\\
WISTELL-A&6.7&4&3\%&1.09&1.03&0.791&2.54\\
ARIES-CS&4.5&3&4\%&0.412&0.498&0.0740&5.69\\
NCSX&4.4&3&4\%&0.392&0.409&0.408&1.55\\
ESTELL&5.3&2&0\%&0.202&0.202&0.570&1.00\\
CFQS&4.3&2&0\%&0.382&0.515&0.586&0.933\\
Henneberg&3.4&2&3\%&0.317&0.314&0.302&2.41\\
QAS2&2.6&2&3\%&0.260&0.267&0.347&1.79\\
\end{tabular}
\caption{\label{tab:configurations}
Quasisymmetric configurations considered in this study.
}
\end{table}

The parameters $B_0$ and $\overline \eta$ are estimated the following way.
We first decompose $B$ into a double Fourier series in the $\vartheta$ and $\varphi$ Boozer angles using the BOOZ\_XFORM code \cite{Sanchez2000a}.
Labelling the poloidal and toroidal Fourier mode numbers by $m$ and $n$ respectively, each Fourier coefficient of $B$ is found as a function of $\psi$, i.e., we write $B_{mn}=B_{mn}(\psi)$.
For each $B_{mn}$, a fifth degree polynomial is fit to the data for $s=\psi/\psi_a$, with $\psi_a$ the toroidal flux at the plasma boundary, in the interval [0, 0.5].
For the $m=0$ modes a polynomial fit in $s$ is used, whereas for the $m=1$ Fourier modes a  polynomial fit in $\sqrt{\psi}$ is used.
For the $m=1$ modes, the data are reflected about $s=0$ to ensure the fit polynomial is odd.
The fitting expression takes the form
\begin{equation}
    B=B_0(\varphi)+r[B_{1c}(\varphi)\cos \vartheta + B_{1s}(\varphi)\sin \vartheta].
\label{eq:B0fit}
\end{equation}
Finally, the parameters $B_0$ and $B_{1c}$ are averaged over $\varphi$, yielding the values in \cref{tab:configurations} for $B_0$ and $\overline \eta$, respectively.

\begin{figure}
    \centering
    \includegraphics[width=0.49\textwidth]{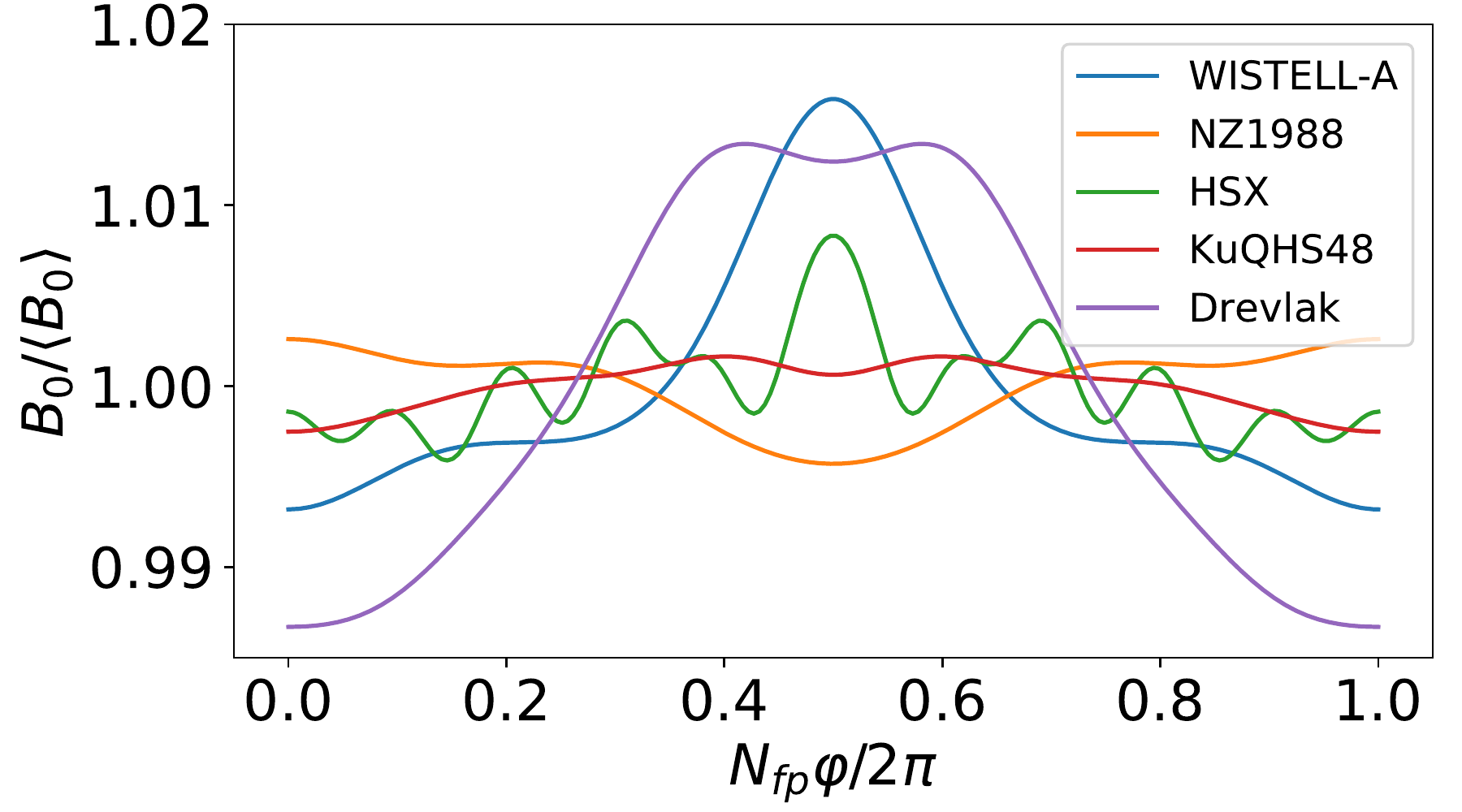}
    \includegraphics[width=0.49\textwidth]{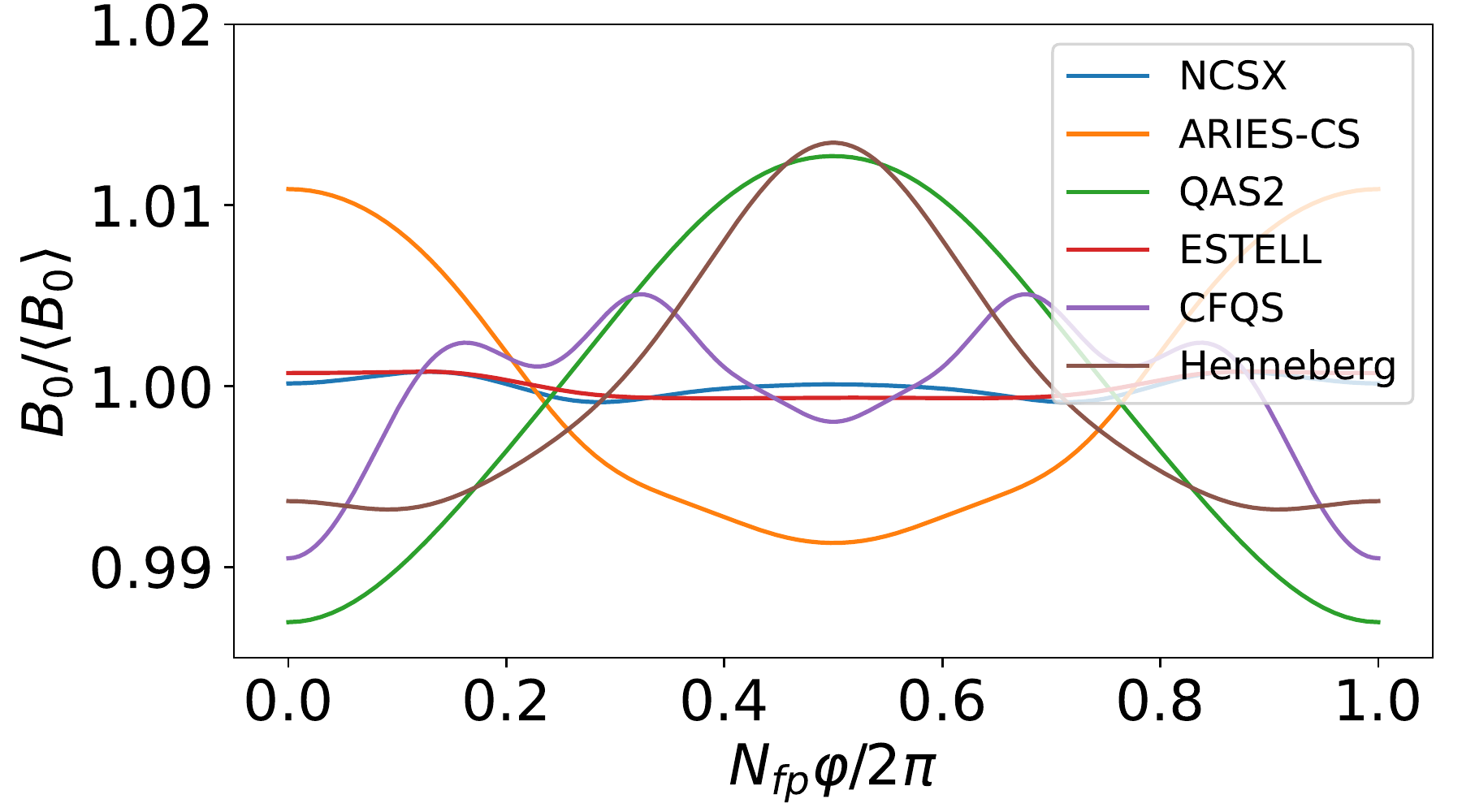}
    \caption{Magnetic field strength on-axis for the quasisymmetric configurations considered on this study normalized by their average value over $\varphi$.}
    \label{fig:B0QS}
\end{figure}

In order to assess the goodness of the fit and how close are the existing configurations to perfect quasisymmetry, we show in \cref{fig:B0QS} the relative variation of magnetic field strength on-axis $B_0$ for the different designs considered in this study.
We expect the agreement to be better (worse) for configurations that have a lower (higher) variation for $B_0$ and that have a higher (lower) aspect ratio.
Most configurations have a relative $B_0$ variation of $\Delta B_0/B_0$ smaller than 2\%.
The configurations with a higher $B_0$ variation are WISTELL-A, Drevlak, ARIES-CS, QAS2 and Henneberg with $\Delta B_0/B_0 \sim 2\%$ while the ones with lower variation are NCSX and ESTELL with $\Delta B_0/B_0 \sim 0.2\%$.
Regarding the Henneberg configuration, we note that this configuration was optimized for quasisymmetry away from the axis, which accounts for its significant $B_0$ variation.
The configurations with lower aspect ratio (therefore higher $\epsilon$ possibly leading to a worse agreement) are QAS2, Henneberg, CFQS and ARIES-CS.
Finally, we mention that of the configurations considered, HSX and CFQS are free-boundary cases with real coils, while the others are fixed-boundary and do not have realistic coil ripple, which explains the fast variation of $B_0$ for HSX in \cref{fig:B0QS} as stemming from coil ripple effects.

\begin{figure}
    \centering
    \includegraphics[width=0.49\textwidth]{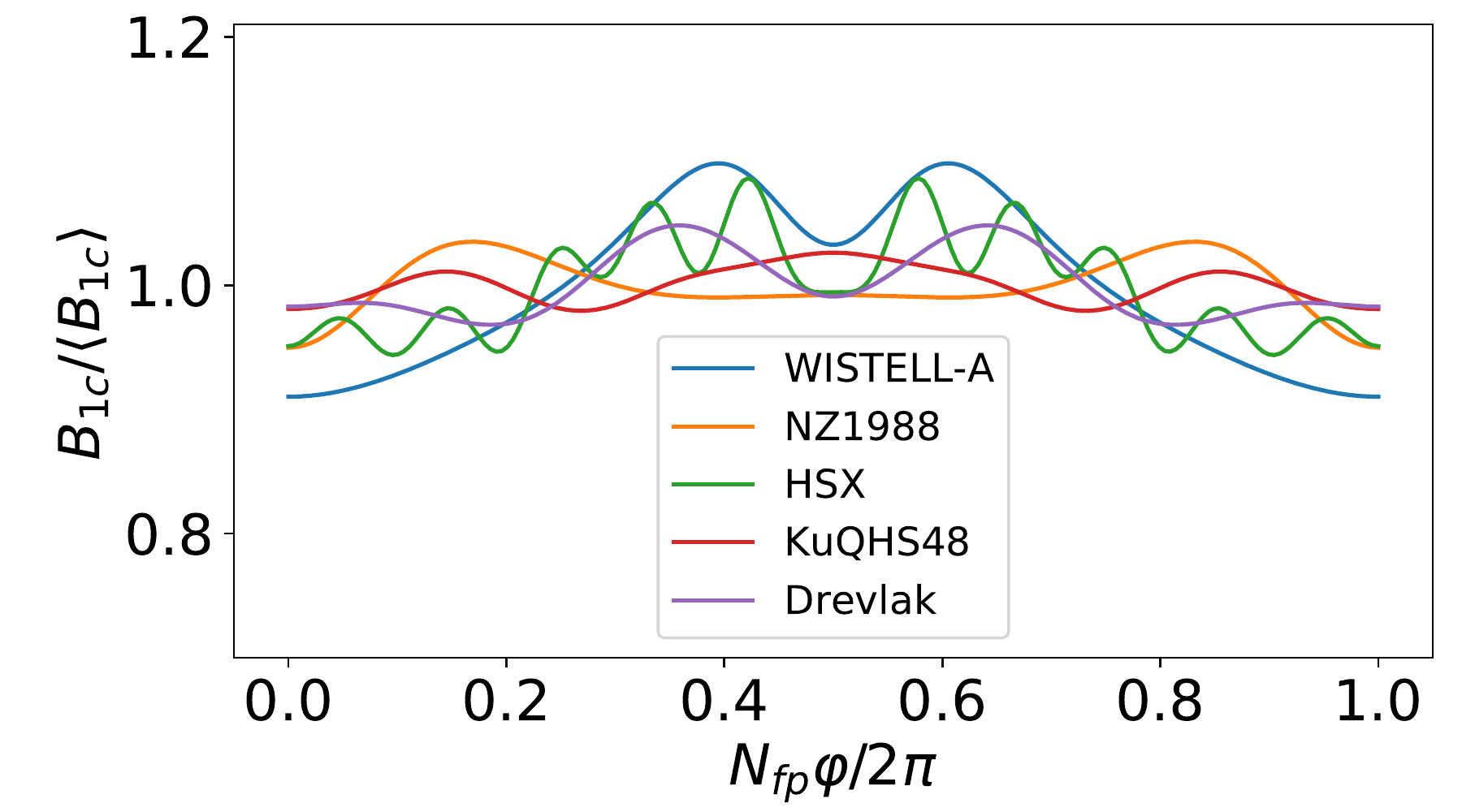}
    \includegraphics[width=0.49\textwidth]{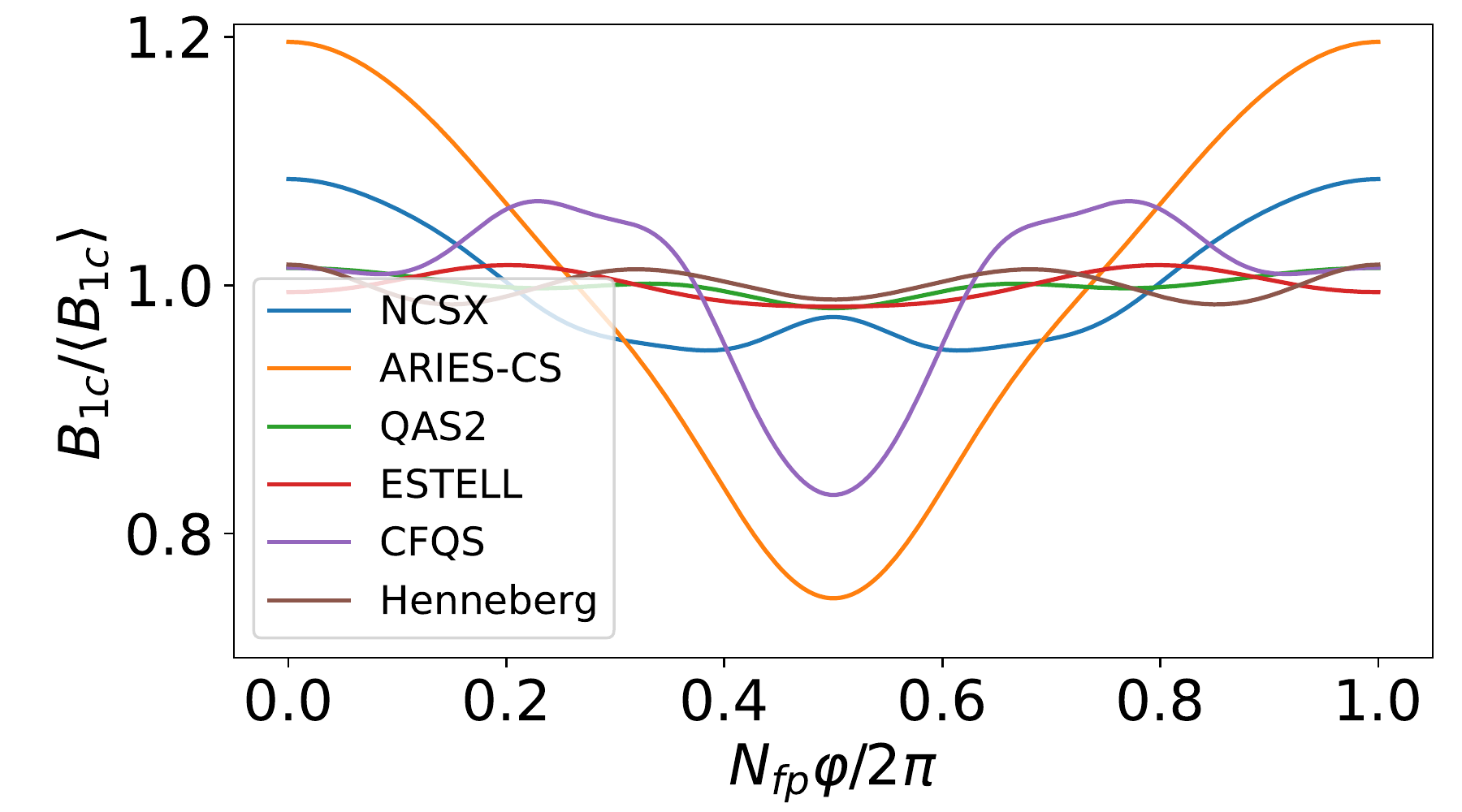}
    \caption{First order magnetic field strength $B_{1c}$ for the quasisymmetric configurations considered on this study normalized by their average value over $\varphi$.}
    \label{fig:B1cQS}
\end{figure}

\begin{figure}
    \centering
    \includegraphics[width=0.49\textwidth]{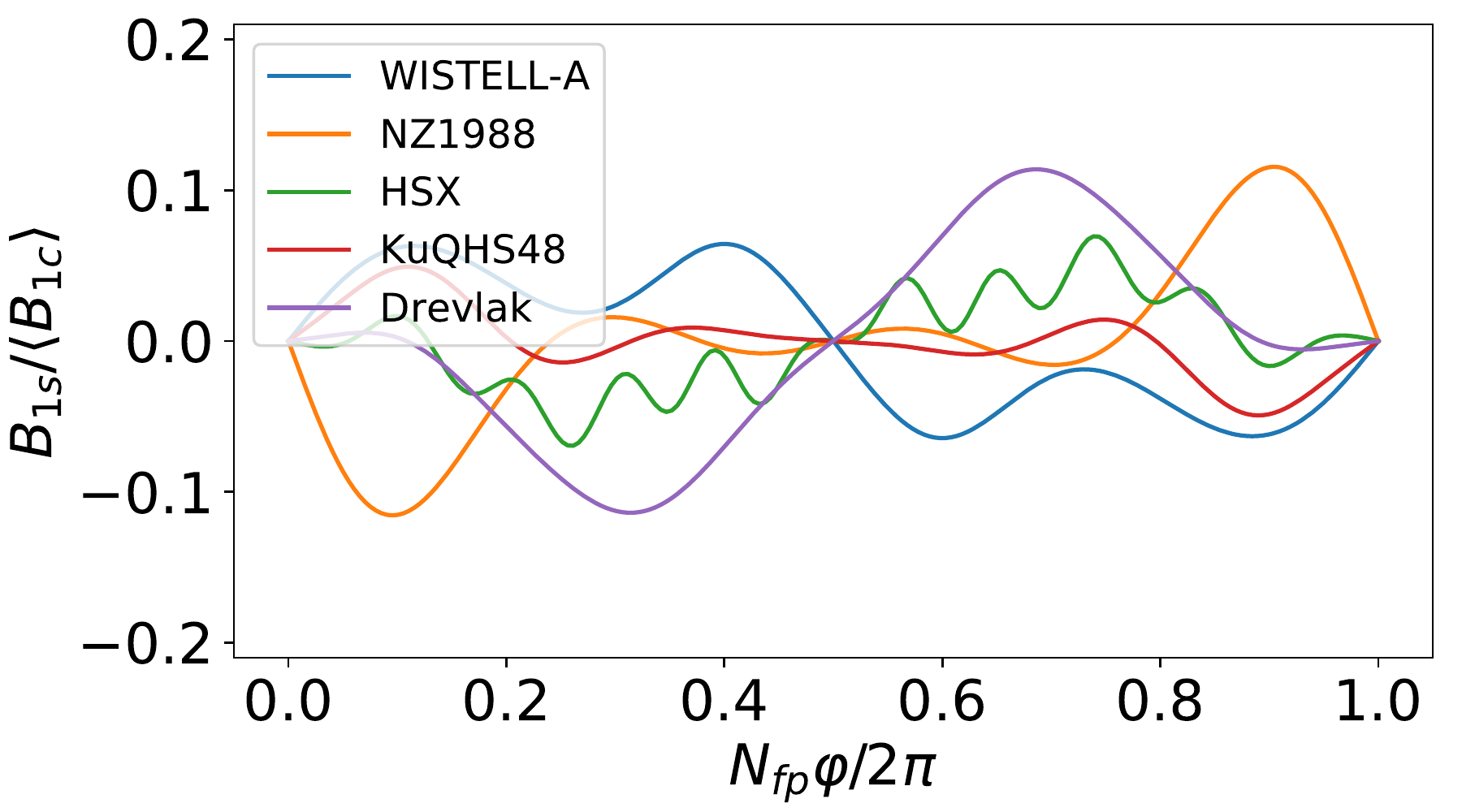}
    \includegraphics[width=0.49\textwidth]{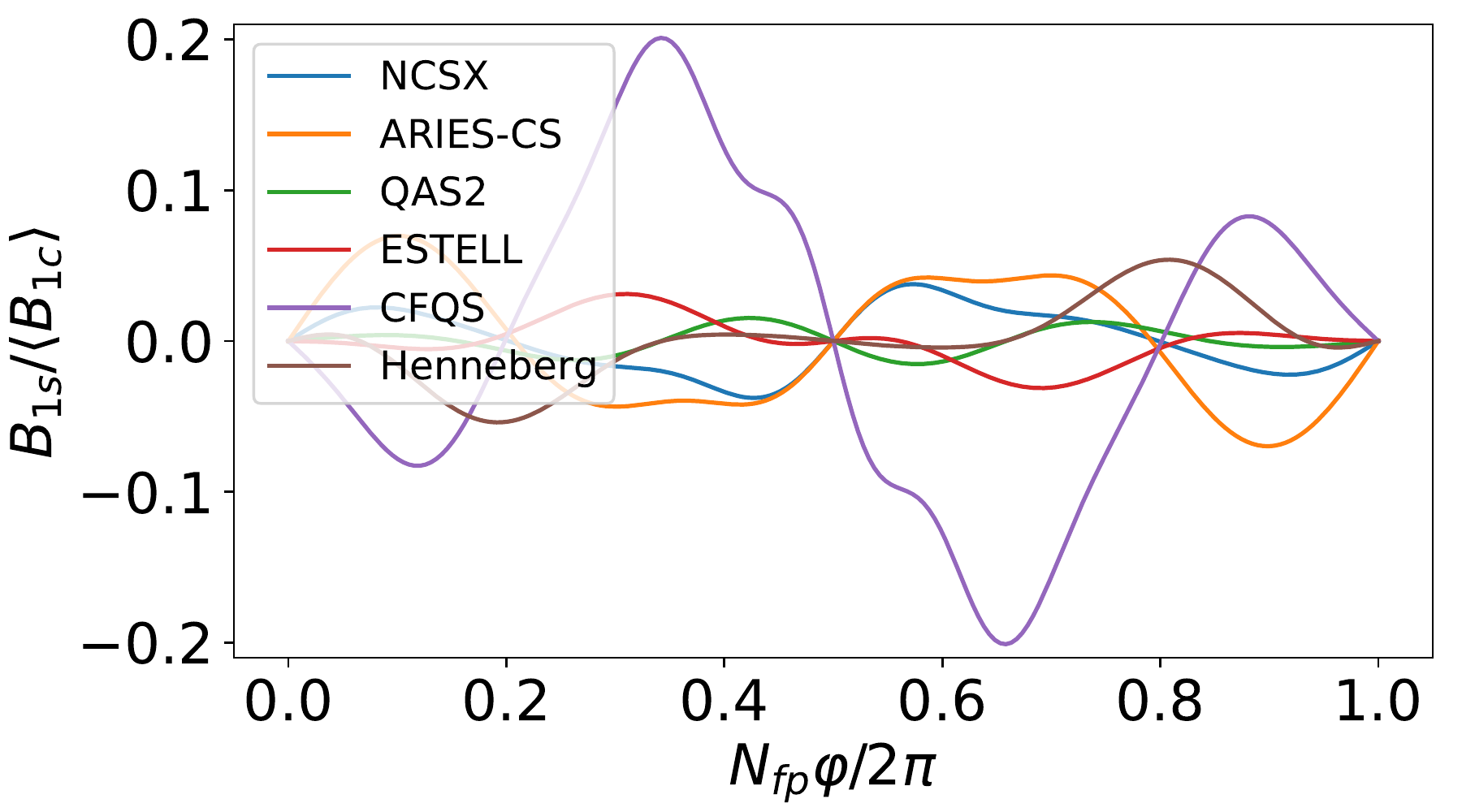}
    \caption{First order magnetic field strength $B_{1s}$ for the quasisymmetric configurations considered on this study normalized by the average value of $B_{1c}$ over $\varphi$.}
    \label{fig:B1sQS}
\end{figure}

An additional source of possible disagreement between the near-axis expansion and the studied configurations is the departure of $B_{1c}$ and $B_{1s}$ from the expected values.
In quasisymmetry, $B_{1c}$ is a constant equal to $B_0 \overline \eta$ while $B_{1s}=0$.
In \cref{fig:B1cQS} we show the function $B_{1c}$ for the configurations considered here, where the largest departure from quasisymmetry is seen for ARIES-CS and CFQS (with $\Delta B_{1c}/B_{1c} \sim 4\%$).
Furthermore, in \cref{fig:B1sQS} we show the variation of $B_{1s}$ from the resulting fit for each of the configurations.
Here, the configuration with the largest variation of $B_{1s}$ is CFQS with $\Delta B_{1s}/B_{1c} \sim 4\%$.

Finally, we provide some additional details on the comparison.
The geometric quantities $\mathbf Q$ of Eq. (\ref{eq:geomQ}) are obtained from the VMEC file \cite{Hirshman1983} corresponding to each design using the geometry module of the \textit{stella} code \cite{Barnes2019}.
As the geometry module uses the cylindrical toroidal angle $\phi$ (not to be confused with the electrostatic potential) as the field line coordinate $z$, in the following, we convert $\varphi$ to $\phi$  using the following result from Ref. \cite{Landreman2018} for quasisymmetric stellarators
\begin{equation}
    \varphi = \frac{2\pi}{L}\int_0^\phi \frac{d\ell}{d\phi} d\phi + O(\epsilon),
\end{equation}
with $d\ell/d\phi$ obtained using $d\ell/d\phi=|\mathbf r_0'(\phi)|$.
Since some elements of $\mathbf{Q}$ diverge or tend to zero on the axis, each element is scaled by the appropriate power of $\psi$ so the on-axis limit is expected to be finite; these scalings are shown in the figure titles.

In order to assess the importance of the departures from quasisymmetry in each configuration, in the comparison involving $\mathbf B \times \nabla B \cdot \nabla \alpha$ and $\mathbf B \times \nabla B \cdot \nabla \psi$, we also show their respective expressions, Eqs. (\ref{eq:g6NA}) and (\ref{eq:g7NA}), where $B_0$ is replaced by the function $B_0(\varphi)$ and, taking Eq. (\ref{eq:g1NA}) into account, both $X_{1c}$ and $X_{1s}$ are replaced by the functions $B_{1c}(\varphi)/\kappa(\varphi) B_0(\varphi)$ and $B_{1s}(\varphi)/\kappa(\varphi) B_0(\varphi)$ found in the fit of Eq. (\ref{eq:B0fit}).
This is denoted as a mixed approach in the comparison figures.
Finally, we note that the comparison will be done in the 'eyeball norm', leaving the assessment of how meaningful the disagreements are using gyrokinetic simulations to future work.

\begin{figure}
    \centering
    \includegraphics[width=0.99\textwidth]{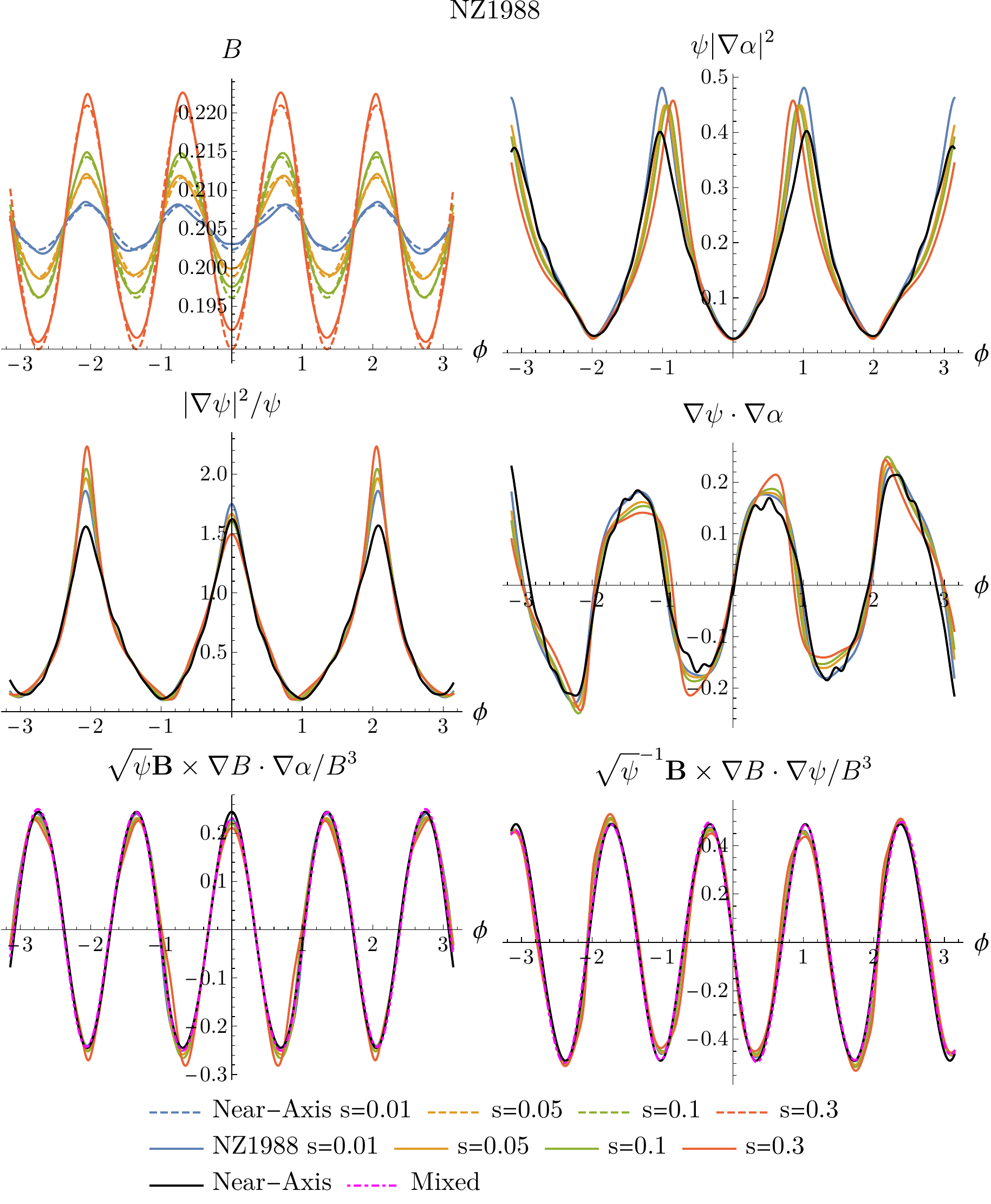}
    \caption{Comparison between the near-axis geometry coefficients and the N\"{u}hrenberg-Zille configuration \cite{Nuhrenberg1988} along the cylindrical toroidal angle $\phi$ at several radial locations $s={\psi/\psi_a}$ with $\psi_a$ the toroidal flux at the plasma boundary. All quantities are expressed in SI units.
    }
    \label{fig:NZ}
\end{figure}

Starting with the comparison in \cref{fig:NZ} for the N\"{u}hrenberg-Zille (NZ1988) configuration, good agreement is obtained between the expansion and the original design even at $s=0.3$, where $s={\psi/\psi_a}$ is the square root of the normalized toroidal flux.
While the match close to the magnetic axis can be attributed to the nearly constant $B_0$ (see \cref{fig:B0QS}), the agreement up to a third of the minor radius is related to the fact that the N\"{u}hrenberg-Zille configuration has a very high aspect ratio, with $A=12$, and low variations of $B_{1c}$ and $B_{1s}$.

\begin{figure}
    \centering
    \includegraphics[width=0.99\textwidth]{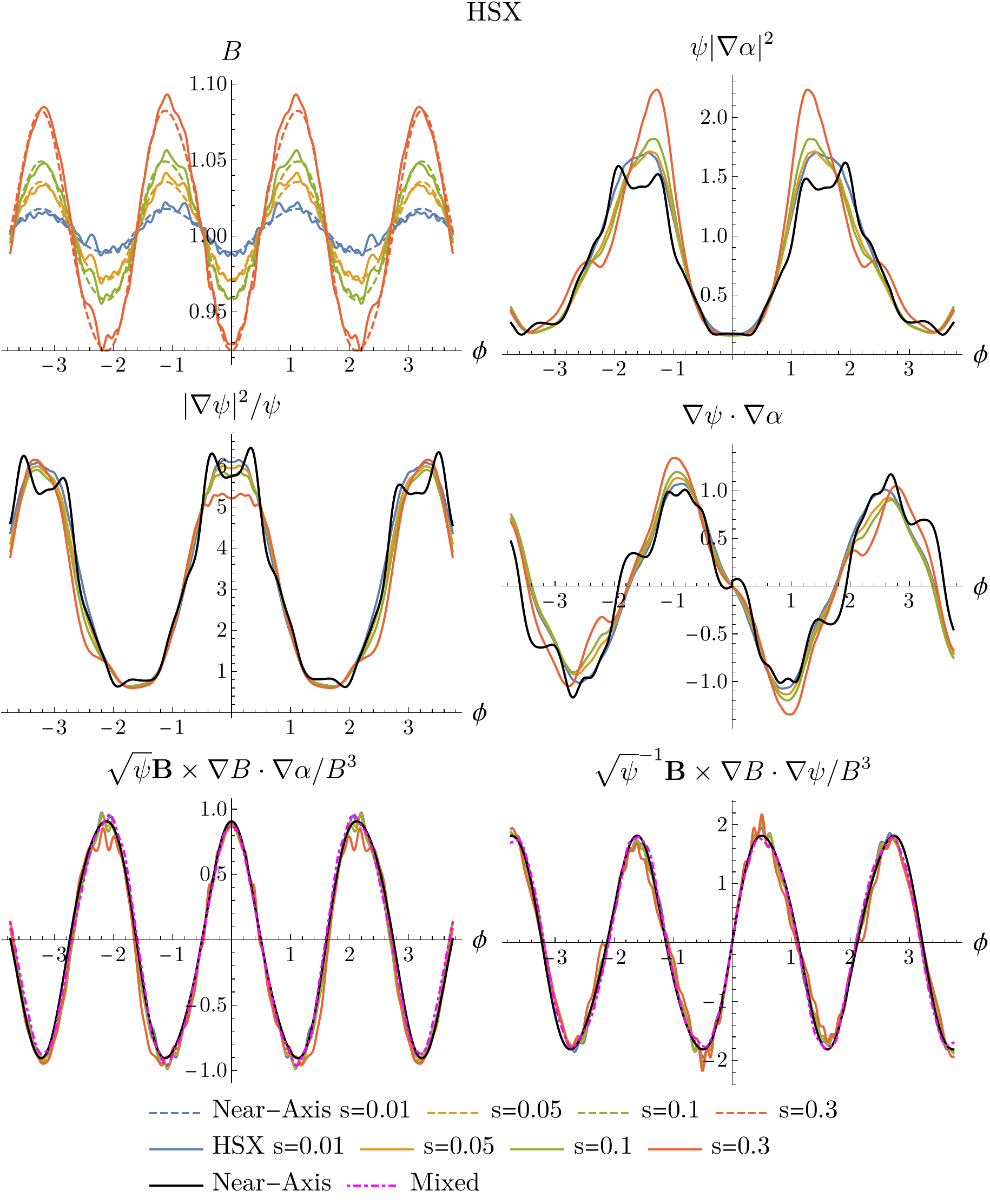}
    \caption{Comparison between the near-axis geometry coefficients and the HSX configuration \cite{Anderson1995} along the cylindrical toroidal angle $\phi$ at several radial locations $s={\psi/\psi_a}$ with $\psi_a$ the toroidal flux at the plasma boundary. All quantities are expressed in SI units.}
    \label{fig:HSX}
\end{figure}

The comparison with the HSX device is shown in \cref{fig:HSX} where we find good agreement for the quantities $B$, $\sqrt{\psi} \mathbf B \times \nabla B \cdot \nabla \alpha/B^3$ and $\sqrt{\psi}^{-1} \mathbf B \times \nabla B \cdot \nabla \psi/B^3$ all the way up to $s=0.3$.
This configuration, similarly to NZ1988, also has a high aspect-ratio, $A=10$, which can explain the agreement.
The remaining geometric quantities, however, show some differences from the prediction of the near-axis expansion, even at $s=0.01$, which can be related to the coil ripple present in \cref{fig:B0QS,fig:B1cQS,fig:B1sQS}.

\begin{figure}
    \centering
    \includegraphics[width=0.99\textwidth]{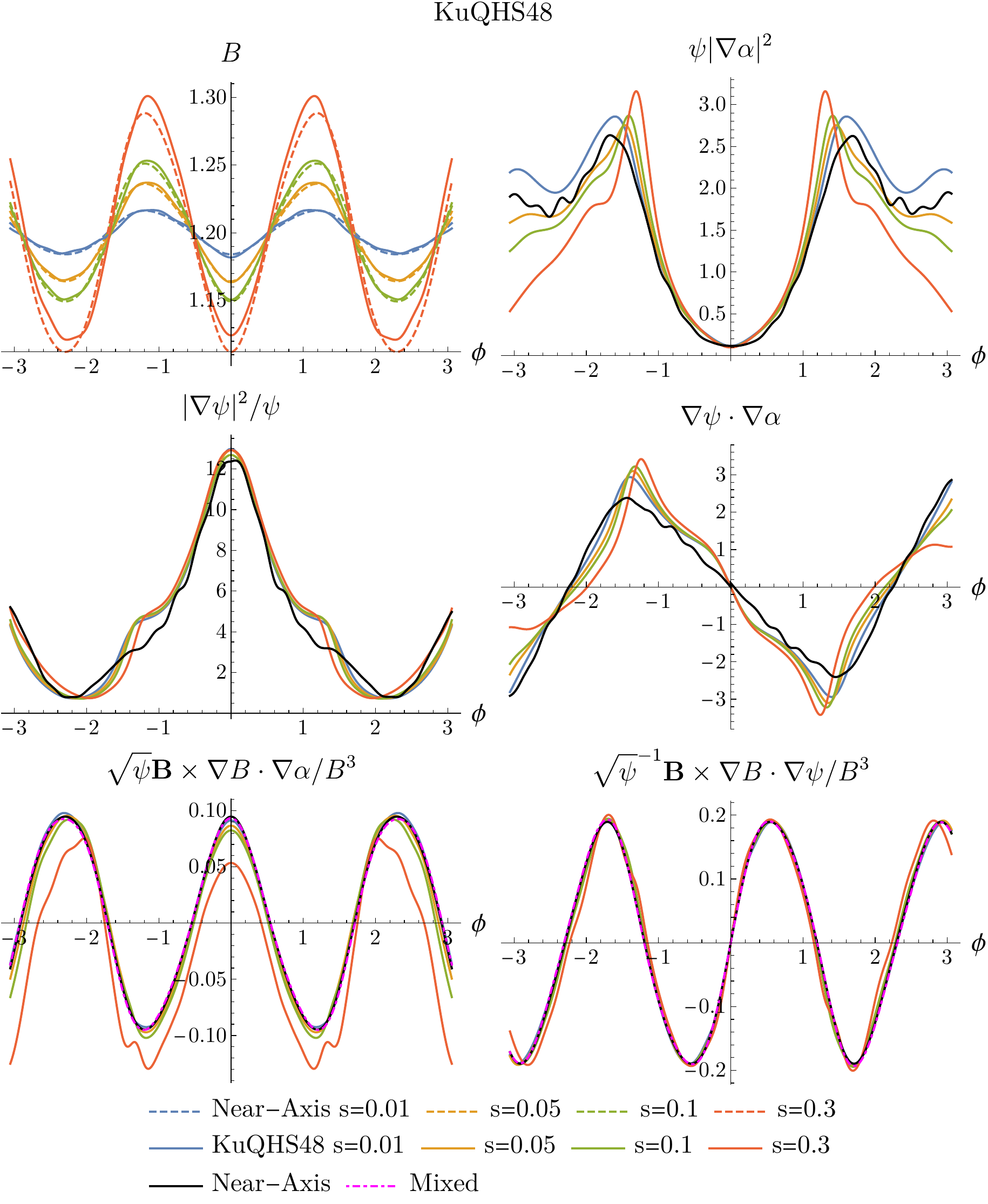}
    \caption{Comparison between the near-axis geometry coefficients and the KuQHS48 configuration \cite{Ku2011} along the cylindrical toroidal angle $\phi$ at several radial locations $s={\psi/\psi_a}$ with $\psi_a$ the toroidal flux at the plasma boundary. All quantities are expressed in SI units.}
    \label{fig:KuQHS48}
\end{figure}

\begin{figure}
    \centering
    \includegraphics[width=0.99\textwidth]{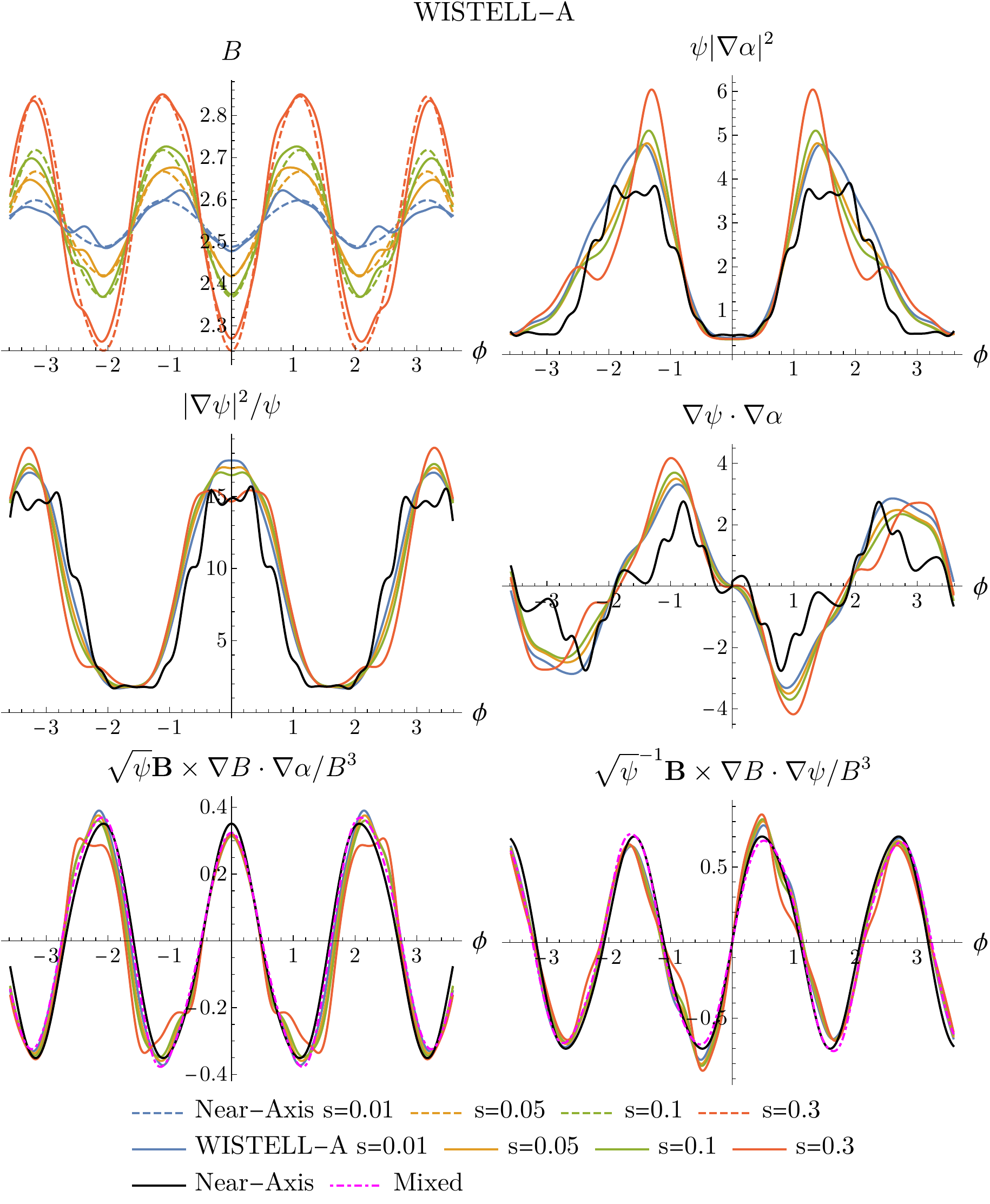}
    \caption{Comparison between the near-axis geometry coefficients and the WISTELL-A configuration \cite{Bader2020} along the cylindrical toroidal angle $\phi$ at several radial locations $s={\psi/\psi_a}$ with $\psi_a$ the toroidal flux at the plasma boundary. All quantities are expressed in SI units.}
    \label{fig:WISTELL-A}
\end{figure}

Next, the KuQHS48 device in \cref{fig:KuQHS48} shows, in general, very good agreement with the prediction from the near-axis framework.
This device has a low variation of $B_0, B_{1c}$ and $B_{1s}$ along $\varphi$ and a high aspect ratio.
A comparison with the WISTELL-A device is shown in \cref{fig:WISTELL-A}.
In general, we find good agreement between the expansion and VMEC except for  $|\nabla \alpha|^2$ and $\nabla \psi \cdot \nabla \alpha$, where the near-axis expansion predicts more short-wavelength structure and lower absolute values of the coefficients, which can be related to the departure from quasisymmetry of the function $B_0$ of WISTELL-A in \cref{fig:B0QS}.
We note that for the profiles of $B$ and $|\nabla \psi|^2$, the agreement is, in fact, better for an of-axis surface with $s=0.3$ than at $s=0.01$.

\begin{figure}
    \centering
    \includegraphics[width=0.99\textwidth]{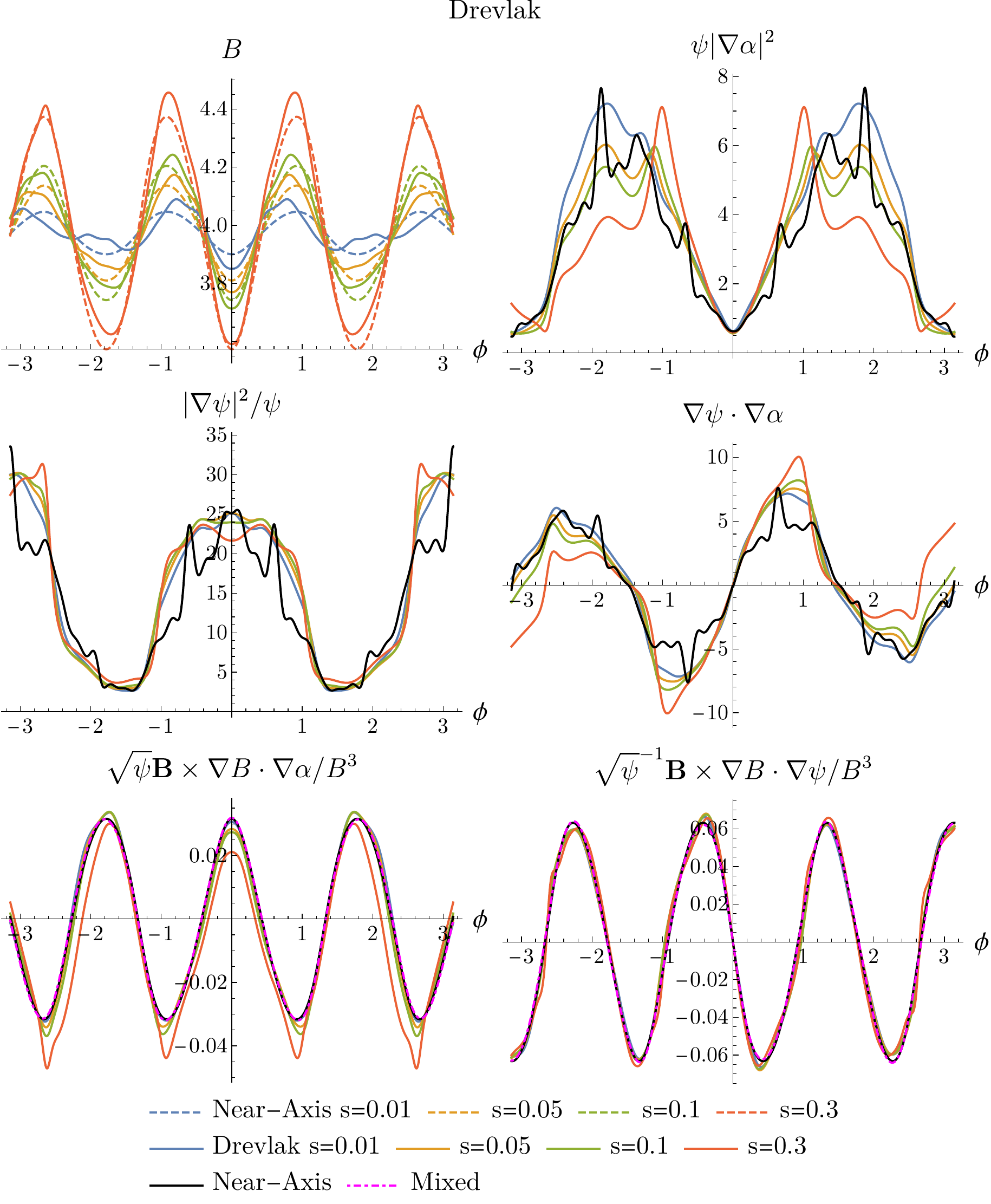}
    \caption{Comparison between the near-axis geometry coefficients and the Drevlak configuration \cite{Drevlak2017} along the cylindrical toroidal angle $\phi$ at several radial locations $s={\psi/\psi_a}$ with $\psi_a$ the toroidal flux at the plasma boundary. All quantities are expressed in SI units.
    }
    \label{fig:Drevlak}
\end{figure}
 
For the Drevlak configuration (\cref{fig:Drevlak}), although the quantities $\sqrt{\psi} \mathbf B \times \nabla B \cdot \nabla \alpha/B^3$ and $\sqrt{\psi}^{-1} \mathbf B \times \nabla B \cdot \nabla \psi/B^3$ are very close to the ones predicted by the near-axis expansion, the magnetic field strength shows variations along the field line with a different behaviour than predicted, even at $s=0.01$.
This may be attributed to the fact that this device, as shown in \cref{fig:B0QS}, has a variation of $B_0$ along $\varphi$ greater than $2\%$.
Such departure from perfect quasisymmetry is similar to the discrepancies observed in $B$ for several values of $s$.
The three remaining coefficients ($|\nabla \psi|^2,|\nabla \alpha|^2$ and $\nabla \psi \cdot \nabla \alpha$), have magnitudes that are similar to the near-axis ones albeit with a different variation along the field line.

\begin{figure}
    \centering
    \includegraphics[width=0.99\textwidth]{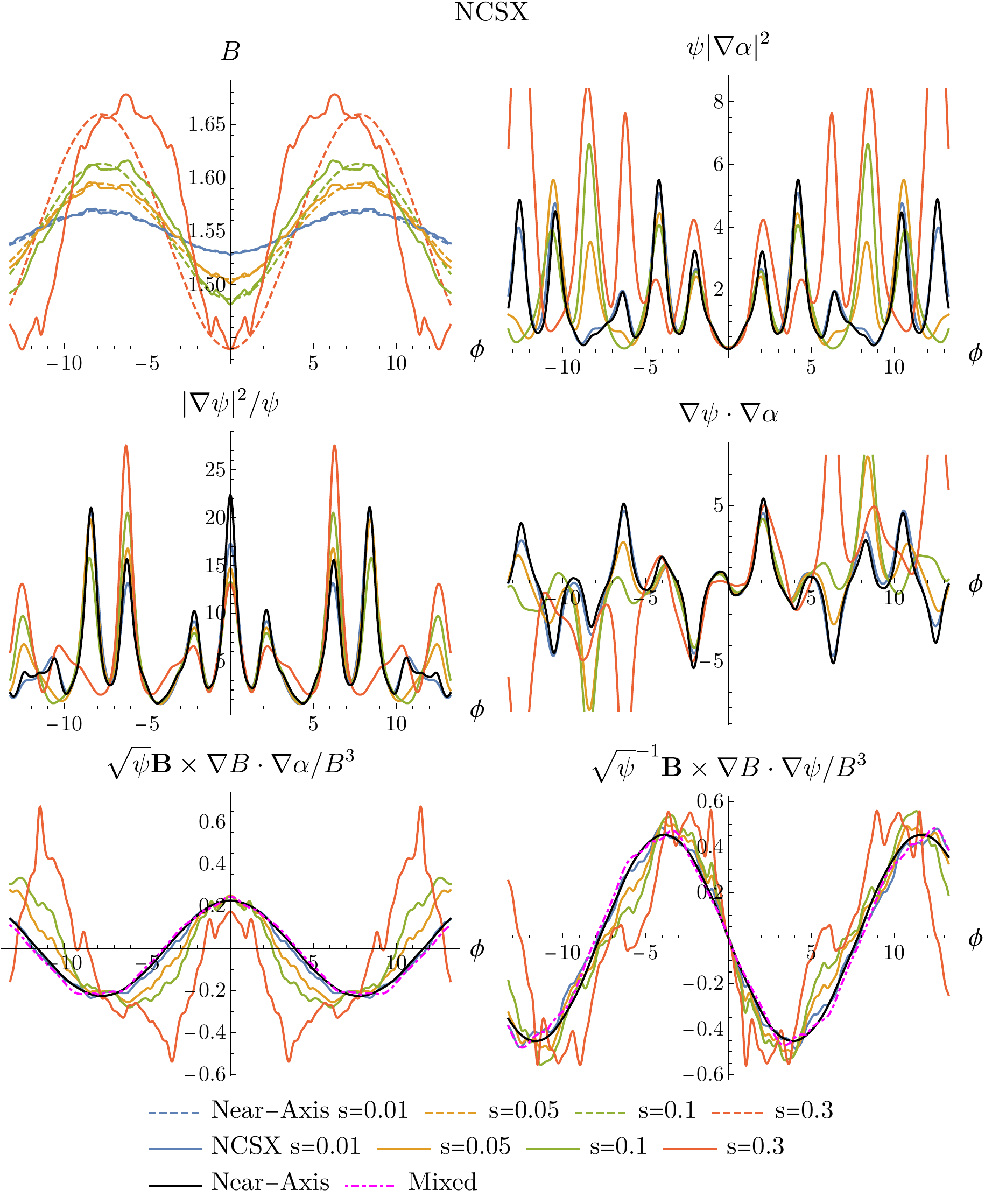}
    \caption{Comparison between the near-axis geometry coefficients and the NCSX configuration \cite{Zarnstorff2001} along the cylindrical toroidal angle $\phi$ at several radial locations $s={\psi/\psi_a}$ with $\psi_a$ the toroidal flux at the plasma boundary. All quantities are expressed in SI units.
    }
    \label{fig:NCSX}
\end{figure}

\begin{figure}
    \centering
    \includegraphics[width=0.99\textwidth]{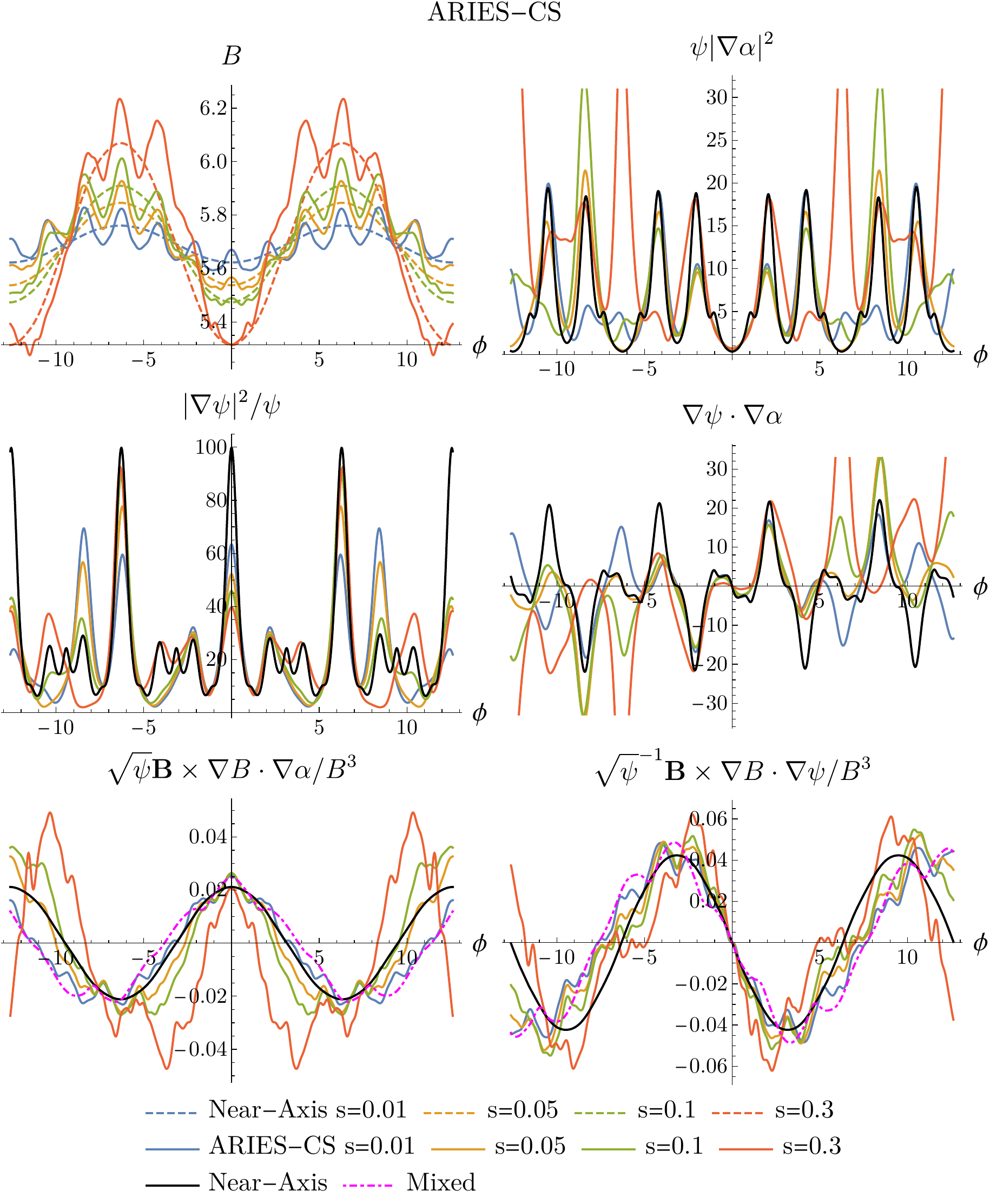}
    \caption{Comparison between the near-axis geometry coefficients and the ARIES-CS configuration \cite{Najmabadi2008} along the cylindrical toroidal angle $\phi$ at several radial locations $s={\psi/\psi_a}$ with $\psi_a$ the toroidal flux at the plasma boundary. All quantities are expressed in SI units.}
    \label{fig:ARIES-CS}
\end{figure}

The comparison with the NCSX device is shown in \cref{fig:NCSX}.
In here, a close match is obtained for $s<0.1$, showing that this device has a very small departure from quasisymmetry on a region close to the magnetic axis.
However, the agreement worsens at $s>0.1$ possibly due to its low aspect ratio of the device, $A=4.4$.
Furthermore, NCSX has a large magnetic shear, which is neglected in the expansion.
This can possibly explain the different wavelengths in $\phi$ between the dashed and solid curves at $s=0.3$ observed in \cref{fig:NCSX}.
A similar situation is found for the ARIES-CS configuration in \cref{fig:ARIES-CS}, which has an aspect ratio similar to the one of NCSX.
However, as ARIES-CS has a larger variation of $B_0$ and, in particular of $B_{1c}$ (see \cref{fig:B0QS,fig:B1cQS}), the geometric coefficients show different variations along the field line than the ones predicted by the expansion.

\begin{figure}
    \centering
    \includegraphics[width=0.99\textwidth]{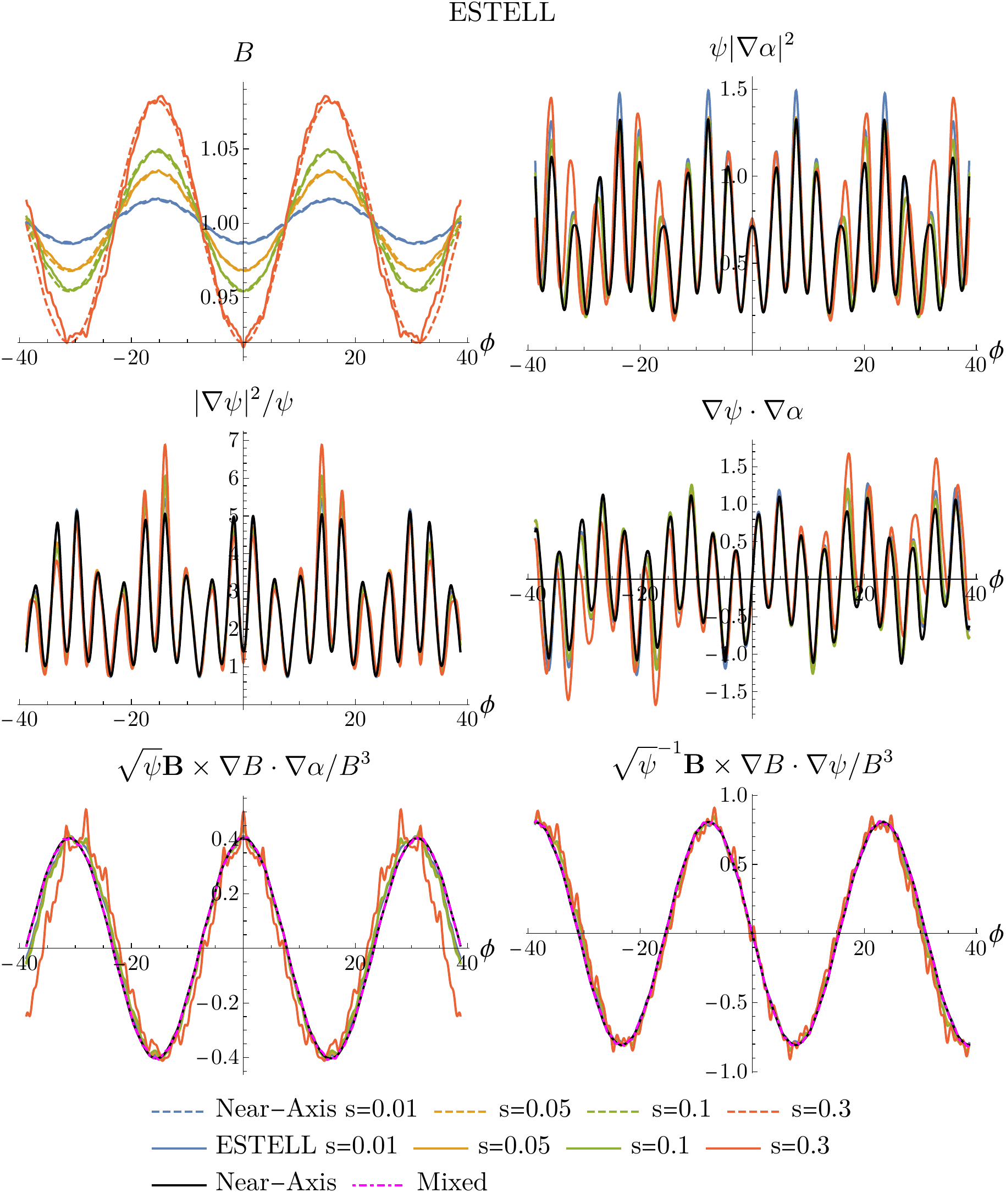}
    \caption{Comparison between the near-axis geometry coefficients and the ESTELL configuration \cite{Garabedian2008} along the cylindrical toroidal angle $\phi$ at several radial locations $s={\psi/\psi_a}$ with $\psi_a$ the toroidal flux at the plasma boundary. All quantities are expressed in SI units.}
    \label{fig:ESTELL}
\end{figure}

\begin{figure}
    \centering
    \includegraphics[width=0.99\textwidth]{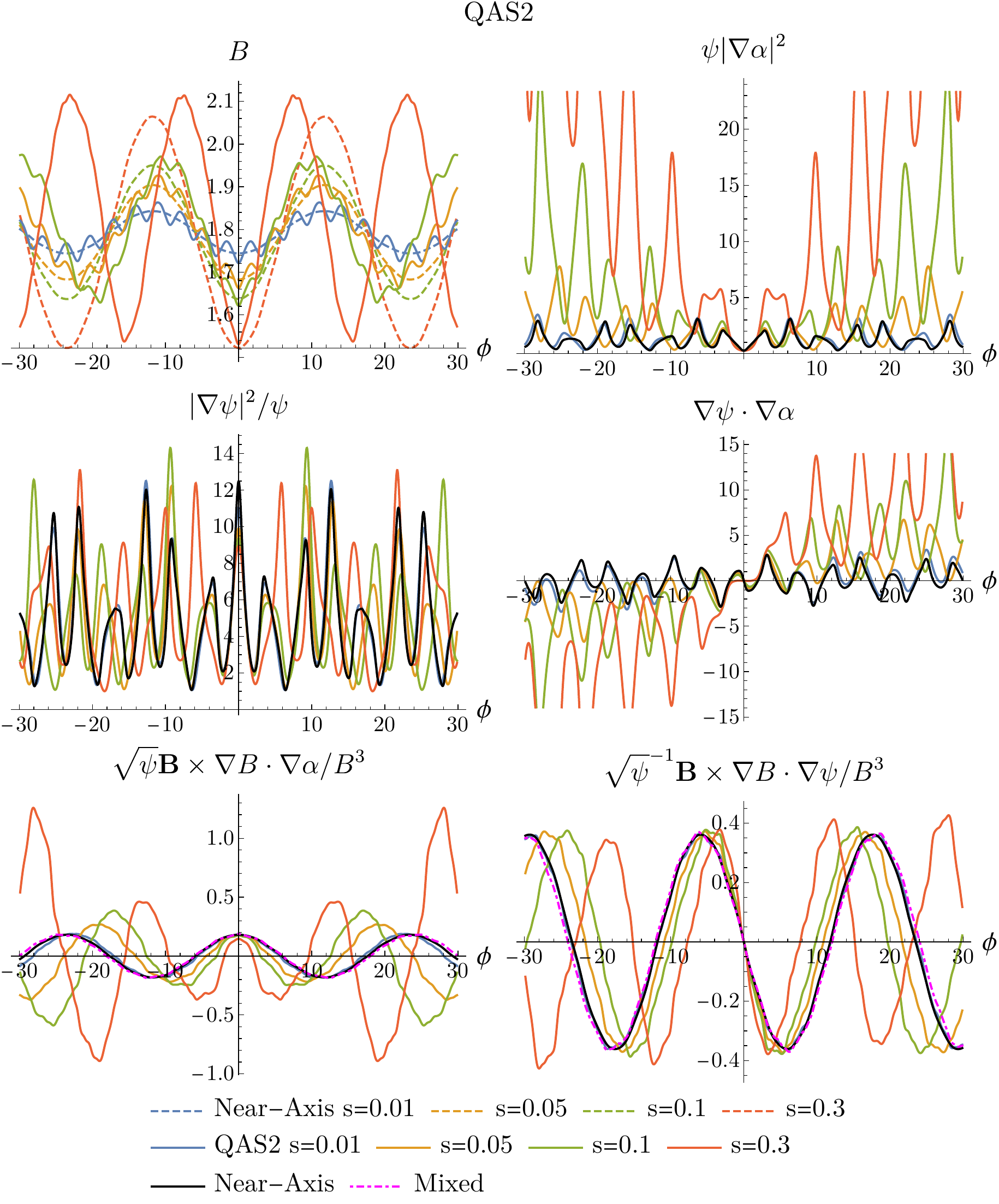}
    \caption{Comparison between the near-axis geometry coefficients and the QAS2 configuration \cite{Drevlak2013} along the cylindrical toroidal angle $\phi$ at several radial locations $s={\psi/\psi_a}$ with $\psi_a$ the toroidal flux at the plasma boundary. All quantities are expressed in SI units.}
    \label{fig:QAS2}
\end{figure}

The ESTELL device (\cref{fig:ESTELL}), which is the device with the overall lowest variation of $B_0, B_{1c}$ and $B_{1s}$, is also the one with the best agreement, which holds even at $s=0.3$.
We note that ESTELL has a low rotational transform and relatively weak shaping when compared with other configurations, which might further explain the observed agreement.
Regarding the QAS2 configuration, this is the device with the lowest aspect ratio and one of the highest variations of $B_0$ in \cref{fig:B0QS}.
Therefore, as seen in (\cref{fig:QAS2}), the agreement with the near-axis prediction worsens substantially as $s$ is increased from $s=0.01$ to $s=0.3$.

\begin{figure}
    \centering
    \includegraphics[width=0.99\textwidth]{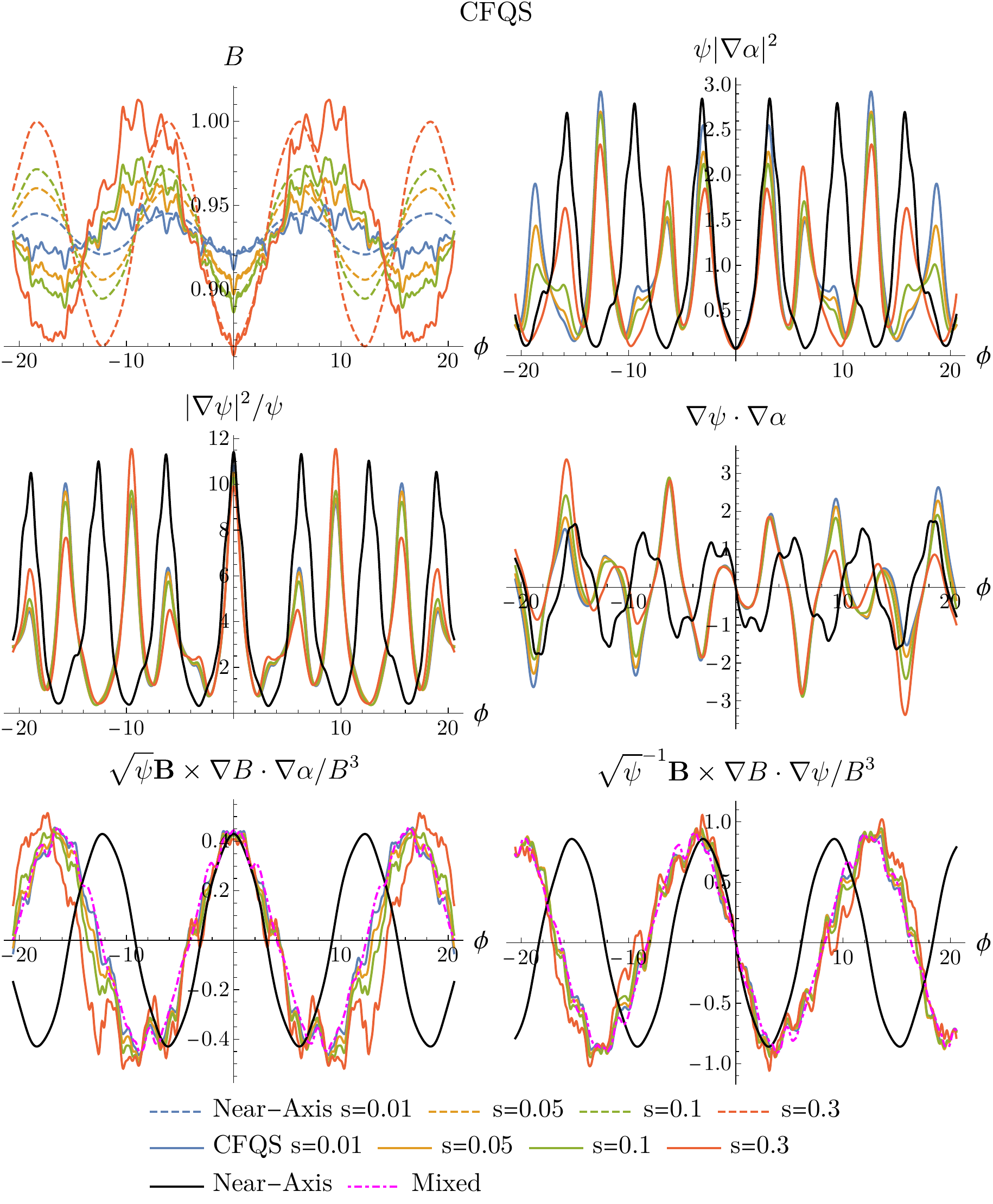}
    \caption{Comparison between the near-axis geometry coefficients and the CFQS configuration \cite{Shimizu2018a} along the cylindrical toroidal angle $\phi$ at several radial locations $s={\psi/\psi_a}$ with $\psi_a$ the toroidal flux at the plasma boundary. All quantities are expressed in SI units.}
    \label{fig:CFQS}
\end{figure}

\begin{figure}
    \centering
    \includegraphics[width=0.99\textwidth]{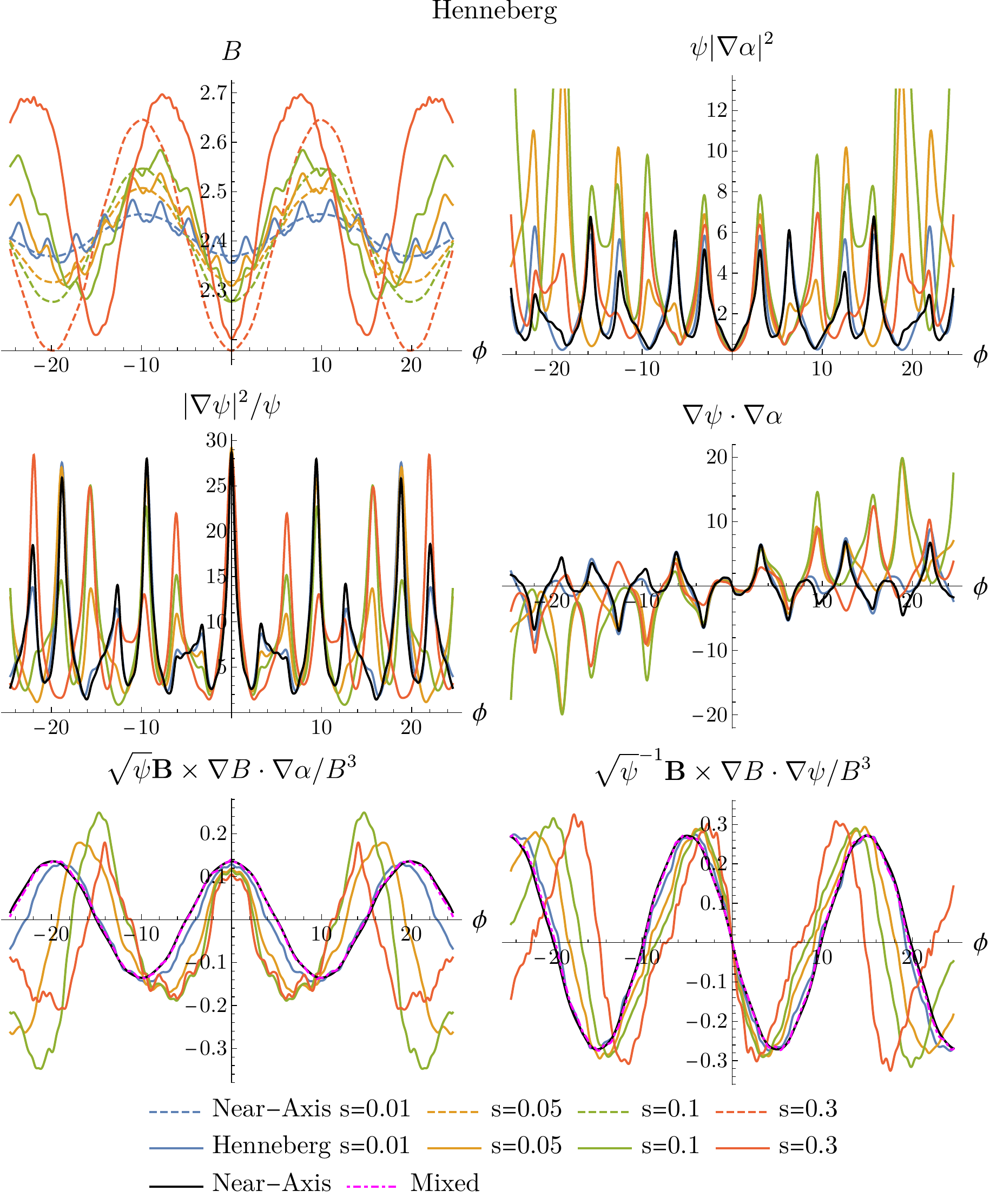}
    \caption{Comparison between the near-axis geometry coefficients and the Henneberg configuration \cite{Henneberg2019} along the cylindrical toroidal angle $\phi$ at several radial locations $s={\psi/\psi_a}$ with $\psi_a$ the toroidal flux at the plasma boundary. All quantities are expressed in SI units.}
    \label{fig:Henneberg}
\end{figure}

The worst agreement with the near-axis prediction is found in \cref{fig:CFQS} for the CFQS device, which is in agreement with the results Ref. \cite{Landreman2019}.
As found there, the rotational transform on axis of the configuration ($\iota=0.382$) differed substantially from the near-axis prediction, $\iota=0.515$. This difference in $\iota$ accounts for the shorter wavelength of the near-axis curves compared to the VMEC curves in each panel of \cref{fig:CFQS}. 
As a possible source of departure from quasisymmetry, we mention the large value of $B_{1s} \sim 0.2 B_{1c}$ observed in \cref{fig:B1sQS}, together with the large variations of $B_0$ and $B_{1c}$ along $\varphi$.
In fact, the mixed approach that takes into account the fact that $B_0$ and $B_{1c}$ have finite variations along $\varphi$, shows a much better agreement for $\sqrt{\psi} \mathbf B \times \nabla B \cdot \nabla \alpha/B^3$ and $\sqrt{\psi}^{-1} \mathbf B \times \nabla B \cdot \nabla \psi/B^3$.
Finally, the Henneberg configuration in \cref{fig:Henneberg} shows good agreement for most quantities up to $s=0.05$, which worsens for higher values of $s$ possibly due to its low aspect ratio of $A=3.4$.
 
As a last comment, we remark on the difference between the dashed and solid curves in the WISTELL-A, Drevlak, QAS2 and Henneberg's configurations.
While the magnetic field strength $B$ of these configurations has short-wavelength ripple that is not predicted by the near-axis expansion, the near-axis approximation for the coefficients $\mathbf B \times \nabla B \cdot \nabla \alpha$ and $\mathbf B \times \nabla B \cdot \nabla \psi$ seems to be very robust.
This pattern can be understood as follows.
For $|B|$, the  expected $\cos(\vartheta)$ variation  is multiplied by a small number $r$ and added to $B_0$ in Eq. (\ref{eq:g1}), so even minor $\varphi$-dependence of $B_0$ causes visible disagreement.
However there is no such addition of a large noisy term in Eqs. (\ref{eq:g6}) and (\ref{eq:g7}) for $\mathbf B \times \nabla B \cdot \nabla \alpha$ and $\mathbf B \times \nabla B \cdot \nabla \psi$, hence these quantities have more robust agreement.

\section{Conclusions}
\label{sec:conclusions}

In the present work, the geometric quantities needed to simulate plasma turbulence in magnetic confinement fusion devices were derived using a near-axis expansion formalism.
These were shown to be identical irrespective if a gyrokinetic, drift-kinetic, two-fluid or ideal ballooning model is employed.
A comparison is made between such geometric quantities obtained using the near-axis formalism and quasisymmetric configurations found in the literature.
Overall, we find good agreement between the two approaches in the  core of the devices.
Possible sources of disagreement include the departure from perfect quasisymmetry in each configuration, the presence of coil ripple, and higher order terms in the expansion. Also, VMEC employs a uniform grid in the radial coordinate $\psi$, which (since $r \propto \sqrt{\psi}$) leads to poor resolution close to the magnetic axis.

We argue that, as the near-axis expansion is based on a framework that can represent the core of stellarator configurations, this method can be used as an effective tool in turbulence optimization studies as it allows for a major reduction of the degrees of freedom involved and the computational time needed for each iteration of the optimization procedure.
While presently the numerical comparison is focused on quasisymmetric magnetic fields, a more general study including non-quasisymmetric fields can be included using the coefficients derived in \cref{sec:generalCase}.
This will be the subject of future studies.
As alternative avenues of future studies, we mention the possibility of using the equations for the near-axis framework to derive the set of geometric coefficients $\mathbf Q$ to next order and improve the matching observed in \cref{sec:comparison}.
Furthermore, we note that gyrokinetic simulations are necessary in order to meaningfully assess whether the differences observed in the comparison are significant or not.

\section{Acknowledgements}
\label{sec:acknowledgements}

This work was supported by a grant from the Simons Foundation (560651, ML) and from the U.S. Department of Energy, Office of Science, Office of Fusion Energy Science, under award number DE-FG02-93ER54197.

\section*{References}
\bibliographystyle{unsrt}
\bibliography{library}

\end{document}